\documentclass[a4paper,11pt]{article}
\pdfoutput=1 
\usepackage{jcappub}

\usepackage{color}
\usepackage{graphicx}
\usepackage{amssymb}
\usepackage[numbers,sort&compress]{natbib}
\usepackage{subcaption}

\newcommand{\calD}{{\ensuremath{\cal D}}}

\newcommand{\dd}{\ensuremath{\textrm{d}}}

\newcommand{\UD}[2]{\ensuremath{^{#1}_{\phantom{#1} #2}}}
\newcommand{\DU}[2]{\ensuremath{_{#1}^{\phantom{#1} #2}}}

\newcommand{\beq}{\begin{equation}}
\newcommand{\eeq}{\end{equation}}
\newcommand{\bea}{\begin{eqnarray}}
\newcommand{\eea}{\end{eqnarray}}
\newcommand{\bit}{\begin{itemize}}
\newcommand{\eit}{\end{itemize}}
\newcommand{\bfi}{\begin{figure}}
\newcommand{\efi}{\end{figure}}
\newcommand{\bfic}{\begin{figure*}}
\newcommand{\efic}{\end{figure*}}
\newcommand{\bce}{\begin{center}}
\newcommand{\ece}{\end{center}}
\newcommand{\bt}{\begin{table}}
\newcommand{\et}{\end{table}}
\newcommand{\btb}{\begin{tabular}}
\newcommand{\etb}{\end{tabular}}

\newcommand{\gzero}{\ensuremath{g^{(0)}}}
\newcommand{\tgzero}{\ensuremath{{\tilde g}^{(0)}}}
\newcommand{\code}[1]{{\texttt{#1}}}

\newcommand{\qed}{\nobreak \ifvmode \relax \else
      \ifdim\lastskip<1.5em \hskip-\lastskip
      \hskip1.5em plus0em minus0.5em \fi \nobreak
      \vrule height0.75em width0.5em depth0.25em\fi}

\title{Light propagation through black-hole lattices}
\author[a,b]{Eloisa Bentivegna,}
\affiliation[a]{
Dipartimento di Fisica e Astronomia\\
Universit{\`a} degli Studi di Catania\\
Via S.~Sofia 64, 95123 Catania\\
Italy}
\affiliation[b]{
INFN\\
Sezione di Catania\\
Via S.~Sofia 64, 95123 Catania\\
Italy}
\emailAdd{eloisa.bentivegna@unict.it}

\author[c]{Miko\l{}aj Korzy\'nski,}
\affiliation[c]{
Center for Theoretical Physics \\
Polish Academy of Sciences \\
Al. Lotnik\'o{}w 32/46, 02-668 Warsaw \\
Poland}
\emailAdd{korzynski@cft.edu.pl}

\author[d]{Ian Hinder,}
\affiliation[d]{
Max-Planck-Institut f\"ur Gravitationsphysik\\
Albert-Einstein-Institut \\ 
Am M\"uhlenberg 1, D-14476 Golm \\
Germany}
\emailAdd{ian.hinder@aei.mpg.de}

\author[d,e]{and Daniel Gerlicher}
\affiliation[e]{
Technische Universit\"at M\"unchen \\
Boltzmannstrasse 15, D-85748 Garching \\
Germany \\
}
\emailAdd{daniel.gerlicher@tum.de}

\abstract{
The apparent properties of distant objects encode information about 
the way the light they emit propagates to an observer, and therefore
about the curvature of the underlying spacetime. Measuring the 
relationship between the redshift $z$ and the luminosity distance $D_{\rm L}$
of a standard candle, for example, yields information on the Universe's
matter content. In practice, however, in order to decode this 
information the observer needs to make an assumption about the 
functional form of the $D_{\rm L}(z)$ relation; in other words, a 
cosmological model needs to be assumed. In this work, we use 
numerical-relativity simulations, equipped with a new
ray-tracing module, to numerically 
obtain this relation for a few black-hole--lattice cosmologies and 
compare it to the well-known Friedmann-Lema{\^i}tre-Robertson-Walker case,
as well as to other relevant cosmologies and to the Empty-Beam Approximation.
We find that the latter provides the best estimate
of the luminosity distance and formulate a simple argument to 
account for this agreement. We also find that a Friedmann-Lema{\^i}tre-Robertson-Walker 
model can reproduce this observable exactly, as long
as a time-dependent cosmological constant is included in the fit. Finally, the dependence
of these results on the lattice mass-to-spacing ratio $\mu$ is discussed: 
we discover that, unlike the expansion rate, the $D_{\rm L}(z)$ relation 
in a black-hole lattice does not tend to that measured in the corresponding 
continuum spacetime as $\mu \to 0$.
}


\begin{document}
\compress
\maketitle

\section{Introduction}
General relativistic spaces filled with black holes have recently 
been under scrutiny as exact cosmological models with a discrete mass
distribution which is, in some sense, uniform on large scales. The
construction of these spaces in numerical relativity has enabled
the investigation of several questions without approximations, such as
how such configurations evolve in time and what their global physical
properties are~\cite{Yoo:2012jz,Bentivegna:2012ei,Yoo:2013yea,Bentivegna:2013jta,Yoo:2014boa}.
At the same time, the numerical simulations have been complemented 
by insight coming from analytical studies, which have illustrated
some general features of these spacetimes such as the behaviour of special 
submanifolds~\cite{Clifton:2013jpa,Clifton:2014lha,Korzynski:2015isa},
the conditions under which they behave like the 
Friedmann-Lema{\^i}tre-Robertson-Walker (FLRW) models~\cite{Korzynski:2013tea},
and the link between their behaviour and the validity of Gauss's law
in a generic theory of gravity~\cite{Fleury:2016tsz}.

In this work, we use numerical spacetimes representing black-hole
lattices (BHLs) to probe a different aspect of inhomogeneous
cosmologies, namely their optical behaviour. 
As is well known, null
geodesics are the bedrock of cosmological observations: light
from distant sources is the primary tool for measuring the 
Universe's density parameters, equation of state, and perturbations. 
Increasing the accuracy of models of light propagation and identifying the 
biases introduced by various approximation frameworks is thus
critical.

Modelling light propagation in inhomogeneous cosmologies is a long-standing
effort, which has followed two complementary courses: approximation 
schemes on one hand, and toy models on the other. The best-known
approach in the former class is the Empty-Beam Approximation (EBA) of
Zeldovich~\cite{1964SvA.....8...13Z}, later generalized by 
Dyer and Roeder~\cite{1972ApJ174L115D,Dyer:1973zz}. This approach
is based on the idea that different effects are at play when light propagates
in a perturbed fluid or through discretely-distributed point masses, as 
different components of the curvature become dominant in either regime
(this is sometimes referred to as the \emph{Ricci-Weyl problem}~\cite{Fleury:2013sna}).
This approach provides an excellent estimate of light propagation in
Swiss-Cheese models, and can be used to constrain the fraction of 
voids in a cosmological model~\cite{Fleury:2013sna,Fleury:2013uqa,Fleury:2014gha}.
The notion that discreteness may affect light propagation more
than inhomogeneity itself has also appeared in other studies, such as 
those on light propagation through Schwarzschild-cell
universes~\cite{Clifton:2009jw,Clifton:2009bp,Clifton:2011mt}.

The existing literature points in a number of common directions:
first, examining individual geodesics, one concludes that the effective 
value of the cosmological constant (the one obtained fitting the 
spacetime to an FLRW model with the same matter density) is higher 
than its microscopic value (the one appearing in the gravitational action).
Second, a statistical average of photon
trajectories usually leads to a partial suppression of this
difference. A suppression is also obtained by considering
the perturbative solution corresponding to a regular arrangements
of objects of equal mass, at least until the perturbative condition
is respected~\cite{Bruneton:2012ru}.

Though consistent on many aspects, these studies are limited
by the conditions imposed on the underlying model: most of the 
discrete-mass studies are either based on spherically-symmetric
building blocks or on the requirements that the objects be not too
compact. It is presently not clear what the optical properties of a more
generic space would be.

To investigate this issue and test the generality of the existing results, 
in this paper we compute the photon redshift and luminosity 
distance along null geodesics running through a BHL spacetime, constructed
exactly through the integration of Einstein's equation, non-perturbatively and 
in three dimensions. First we compare
the result to some reference models from the FLRW class, to the Milne
cosmology, and to a generic universe in the Empty Beam Approximation (EBA)~\cite{1964SvA.....8...13Z,1972ApJ174L115D,Dyer:1973zz}.
We find that the latter provides the closest approximation to light
propagation on the BHL, and derive a simple argument to explain this result,
which in some sense extends the reasoning of~\cite{Fleury:2014gha}
to completely vacuum spacetimes.
We then turn to the question of whether it is possible to tune the
cosmological parameters in the FLRW class to improve the fit. We find, in 
particular, that one can reproduce the luminosity-distance--to--redshift
relationship of a BHL with that of an FLRW model with the same average
matter density and a fictitious, time-dependent
cosmological constant $\Lambda$, and provide the first measurement of
this running in our base configuration. Finally, we study how this behaviour 
depends on the BHL inhomogeneity parameter $\mu$~\cite{Bentivegna:2013jta},
which roughly corresponds to the ratio between the central mass and the 
lattice spacing, and in particular we analyse the continuum limit of $\mu \to 0$.

An important factor in this discussion is the choice of light 
sources and observers, as the photon frequencies and number counts
will depend on the reference frame in which they are measured. 
In FLRW models there is an obvious option: the comoving sources and 
observers. In inhomogeneous spaces, on the other hand, identifying a 
``cosmic flow'' is more tricky (when at all possible) and relies on the 
somewhat arbitrary split between global cosmological evolution and 
``local effects'' sourced by nearby gravitational structures.
For the purpose of this work, we sidestep this question
by noticing that, for a given geodesic, the angular and luminosity distances
can be obtained by applying a certain linear operator to the four-velocity
of the observer, with no dependence whatsoever on the motion of the light source.
It is therefore straightforward to quantify the effect of different
observer prescriptions on these observables.

Section~\ref{sec:lprop} introduces the formalism of light propagation and justifies
the approach we take in our analysis, providing some examples
in simple spacetimes. Section~\ref{sec:bhl} provides an approximate description
of light propagation in a BHL via a perturbative analysis.
We present the numerical results in section~\ref{sec:results} and in section~\ref{sec:dc} we comment on them.
We provide tests of the geodesic integrator, used for the first time
in this study, in the appendix.
We use geometric units $G=c=1$ everywhere.

\section{Fundamentals of light propagation}
\label{sec:lprop}
Let us start by considering a null ray emanating from a light
source $\cal{S}$ and reaching an observer $\cal{O}$: this curve can be 
described as an affinely-parametrized null geodesic $\gamma(\lambda)$, with 
$\cal{S}$ and $\cal{O}$ as end points corresponding to the affine parameter
values $\lambda_{\cal{S}}$ and $\lambda_{\cal{O}}$:
\bea
\gamma(\lambda_{\cal{S}}) = \cal{S}\\
\gamma(\lambda_{\cal{O}}) = \cal{O}
\eea
The curve is described by the geodesic equation:
\beq
\label{eq:GE}
{ \nabla_p} p^a = 0
\eeq
where:
\beq
p^a = \frac{\dd x^a}{\dd \lambda}
\eeq
is the tangent vector to $\gamma$.
In order to measure distances with null rays, however, we need more
than a single geodesic: we need to consider a whole {\it beam} of 
rays~\cite{Seitz:1994xf}, centred on $\gamma$, and study the evolution 
of its cross-sectional area as it makes its way from $\cal{S}$ to 
$\cal{O}$. 

The time evolution of a beam's cross section is described by the 
geodesic deviation equation (GDE). Let $\xi^a$ be the separation 
vector between the fiducial geodesic $\gamma$ and a neighbouring 
one, called $\tilde \gamma$. It satisfies 
\bea
 \nabla_p \nabla_p \xi^a = R\UD{a}{bcd}\,p^b\,p^c\,\xi^d. \label{eq:GDE}
\eea
The GDE is a second order ODE for the 4--vector $\xi^a$, or equivalently 
a first order ODE for $\xi^a$ and $\nabla_p \xi^a$.
It is valid for any neighbouring geodesic, but since in the geometrical optics 
we are only interested in null geodesics, we impose a restriction 
on the solution $\xi^a(\lambda)$ of the form:
\bea
 p_a \nabla_p \xi^a = 0, \label{eq:null}
\eea
which ensures that $\tilde\gamma$ is null.
Note that if the equation above is satisfied at one point, then it is 
automatically satisfied along the whole of $\gamma$ because of equation 
(\ref{eq:GDE}).

Let us now restrict the geodesics under consideration to those which 
lie on the same wavefront as $\gamma$, i.e.~for which the separation vector satisfies
\bea
 \xi^a\,p_a = 0. \label{eq:wavefront}
\eea
The condition above means that, for a given observer at a given time, the 
photon corresponding to the geodesic $\gamma$ and the one corresponding to
$\tilde \gamma$ lie on the same 2-plane perpendicular to the direction of 
propagation (see Figure~\ref{fig:wavefronts}). This condition is Lorentz-invariant, meaning that 
if it is satisfied in one reference frame then it is valid in all frames. 
Moreover, for null geodesics it propagates along $\gamma$, i.e.~if it is 
satisfied at one time it is satisfied along the whole of $\gamma$. This 
follows easily from (\ref{eq:null}) and (\ref{eq:GDE}).

\bfi[!h]
\centering
\includegraphics[width=0.5\textwidth]{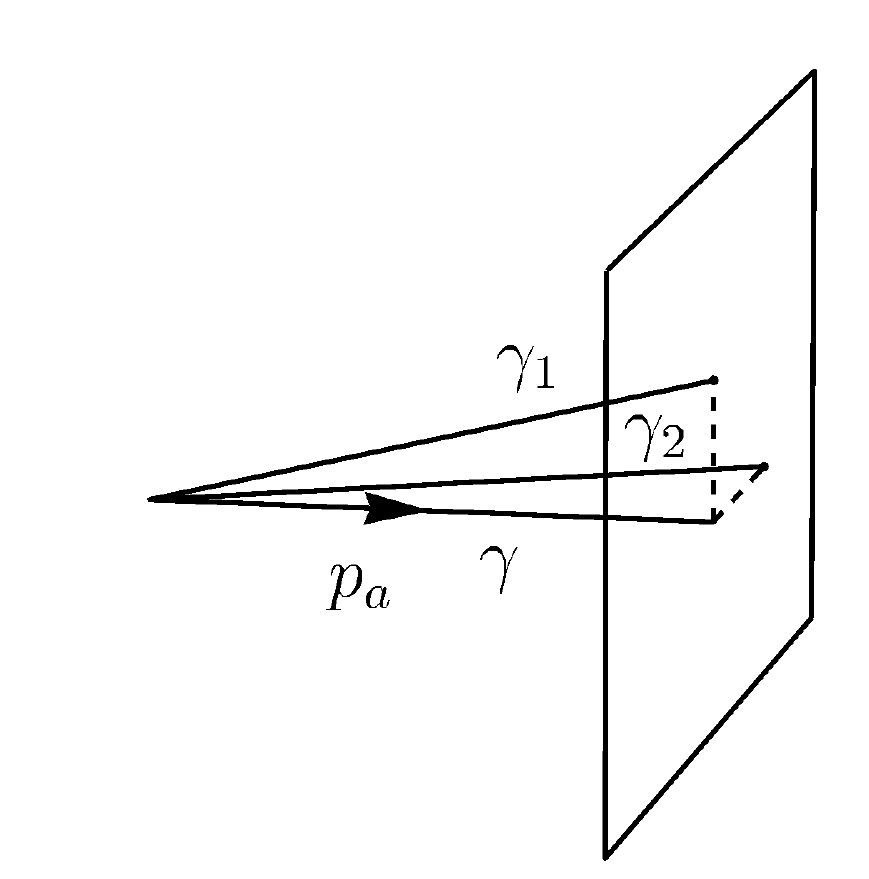}
\caption{The null geodesics lying on the same wavefront consist of geodesics for which the photons at any instant of time
and for any observer lie on the same plane perpendicular to the direction of propagation given by $p_a$.
\label{fig:wavefronts}}
\efi

The reason why we are interested only in geodesics which lie on the same 
wavefront is that we want to study geodesics which cross at one point, 
either the emission point ${\cal S}$ or the observation point ${\cal O}$. 
If this is the case, then $\xi^a = 0$ at either $\lambda_{\cal O}$ or 
$\lambda_{\cal S}$, so that (\ref{eq:wavefront}) is trivially satisfied there 
and thus also \emph{everywhere} on $\gamma$.

By imposing (\ref{eq:null}) and (\ref{eq:wavefront}) we have effectively 
reduced the number of degrees of freedom from four to three. It turns out 
that a further reduction is possible.
Note that at every point we are free to add a vector proportional to  $p^a$ 
to both $\xi^a$ and $\nabla_p\xi^a$. The former corresponds to using a 
different point of \emph{the same} geodesic $\gamma$ in the definition of 
the separation vector $\xi^a$, while the latter is just a rescaling of the 
affine parametrization of $\gamma$. Neither transformation affects the 
physical content of the equations, as long as we are in the regime of 
geometrical optics. As a matter of fact, it is easy to see that equations 
(\ref{eq:GDE})--(\ref{eq:wavefront}) are insensitive to these transformations 
as well:
\bea
&&\nabla_p \nabla_p \left(\xi^a + C(\lambda)\,p^a\right) = R\UD{a}{bcd}\,p^b\,p^c\,\xi^d + \ddot C\,p^a \\
&&\nabla_p \left(\xi^a + C(\lambda)\,p^a\right)\,p_a = \dot C\,p^a\,p_a = 0 \\
&&\left(\xi^a + C(\lambda)\,p^a\right)\,p_a = C\,p^a\,p_a = 0. 
\eea
It follows that (\ref{eq:GDE})--(\ref{eq:wavefront}) can be reinterpreted as 
equations on the space $p^\perp/p$, consisting of vectors orthogonal to 
$p_a$ and divided by the relation $\xi^a \sim \eta^a \iff \xi^a = \eta^a + A\,p^a$. 
We shall denote the equivalence class corresponding 
to a vector $\xi^a$ in $p^\perp$ as $\left[\xi\right]^A$. 
The space $p^\perp/p$ is two--dimensional and inherits the positive-definite 
metric from $g_{ab}$ via the relation $\left[X\right]^A\,\left[Y\right]^B\,g_{AB} 
= X^a\,Y^b\,g_{ab}$, where $X^a$ and $Y^b$ are any vectors in the tangent 
space corresponding to the equivalence classes $[X]^A$ and $[Y]^B$, respectively.
It can be thought of as the space of null geodesics lying in the neighbourhood 
of $\gamma$ on the same wavefront, without any specification of which point on $\gamma$ we assign to 
which point of $\tilde \gamma$. It is straightforward to verify that the covariant derivative
$\nabla_p$ can also be defined as an operator on $p^\perp/p$.

In the standard formalism due to Sachs \cite{Sachs309, lrr-2004-9}, we then 
introduce a frame with two spatial, orthonormal screen vectors $\xi_1^a$ and $\xi_2^a$, 
both orthogonal to $p^a$ and to a timelike observer $u^a_{\cal O}$. Notice that
this is not strictly necessary:  
all that matters in geometrical optics are the \emph{equivalence classes} 
$\left[\xi_1\right]^A$ and $\left[\xi_2\right]^B$, which turn out to be entirely 
$u^a_{\cal O}$\emph{-independent}. More precisely, for any other choice of the observer 
$\tilde u^a_{\cal O}$ and the corresponding $\tilde \xi_1^a$ and $\tilde \xi_2^a$ perpendicular to 
$p_a$, the classes $\left[\tilde \xi_1\right]^A$ and $\left[\tilde \xi_2\right]^B$ are 
related to $\left[\xi_1\right]^A$ and $\left[\xi_2\right]^B$ via a simple spatial rotation.

The image distortion of a distant object and its angular distance can now be 
calculated by finding the Jacobi matrix $\calD\UD{A}{B}$ of the GDE in the space $p^\perp/p$
\bea
\nabla_p \nabla_p \calD\UD{A}{B} = R\UD{A}{\mu\nu C}\,p^\mu\,p^\nu\,\calD\UD{C}{B} \label{eq:GDE2}
\eea
with the initial data of the form
\bea
&&\calD\UD{A}{B}(\lambda_{\cal O}) = 0 \label{eq:ID2} \\
&&\nabla_p \calD\UD{A}{B}(\lambda_{\cal O}) = \delta\UD{A}{B} \nonumber
\eea
(see \cite{lrr-2004-9} for its geometric definition and the discussion of its properties). Note that the initial data depends on the choice of parametrization of the null 
geodesic $\gamma$, because if we rescale $\lambda \mapsto C\,\lambda$, the tangent 
vector rescales accordingly via $p^a \to C^{-1} p^a$. Thus $\calD\UD{A}{B}$ is 
parametrization-dependent. Nevertheless, the tensor product $p_\mu\,\calD\UD{A}{B}$ 
is parametrization-independent and is therefore an intrinsic property of the light cone 
centred at the observation point $\cal O$. In practice the equations (\ref{eq:GDE2})--(\ref{eq:ID2})
are solved by first introducing a Sachs frame and then using the corresponding 
screen vectors $\left[ \xi_1\right]^A$ and $\left[ \xi_2\right]^B$ as a basis in $p^\perp/p$. 

The image distortion seen by the observer with 4-velocity $u_{\cal O}^a$ at the 
observation point is finally:
\bea
I\UD{A}{B} = \left|u_{\cal O}^a\,p_a\right|\,\calD\UD{A}{B}(\lambda_{\cal S})
\eea
while the angular distance is
\bea
\label{eq:DA}
D_{\rm A}=\left|u_{\cal O}^a\,p_a\right|\,\sqrt{\left|\det \calD\UD{A}{B}(\lambda_{\cal S})\right|}
\eea
 (see also \cite{lrr-2004-9} and references therein). Note that the result does not depend on the 
 4-velocity of the source, while the dependence on the 4-velocity of the observer is quite simple.
For instance, it is easy to prove that, on an FLRW spacetime, observers boosted with respect to the
comoving frame measure smaller angular distances, because the quantity $\left|u_{\cal O}^a\,p_a\right|$
decreases as the boost parameter is increased. One can therefore use equation (\ref{eq:DA}) to work
out which observers (if any) would measure a specified angular distance for an object in a given 
spacetime. 

 The luminosity distance is defined using the total energy flux from the source through a fixed area at the observation point. In the formalism above
 it can be expressed as
 \bea
 \label{eq:DL}
  D_{\rm L}=\left|u_{\cal S}^a\,p_a\right|\,\sqrt{\left|\det \tilde\calD\UD{A}{B}(\lambda_{\cal O})\right|}(1+z)
 \eea
 where $\tilde\calD\UD{A}{B}$ satisfies (\ref{eq:GDE2}), but with the initial conditions (\ref{eq:ID2}) imposed at the source rather than at the observer,
 and $z$ is the relative change in the photon frequency as it moves along the geodesic, also known as its {\it redshift}:
 \beq
  z = \frac{\nu_{\cal S}-\nu_{\cal O}}{\nu_{\cal O}} = \frac{u_{\cal S}^a\,p_a}{u_{\cal O}^a\,p_a} - 1.
 \eeq
 The fundamental result by Etherington \cite{springerlink:10.1007/s10714-007-0447-x} relates these quantities: the reciprocity relation reads
 \bea
 \left|\det \tilde\calD\UD{A}{B}(\lambda_{\cal O})\right| =  \left|\det \calD\UD{A}{B}(\lambda_{\cal S})\right|. \label{eq:Etherington}
 \eea
 It follows easily that
 \beq
 D_{\rm L} = (1+z)^2 D_{\rm A}.
 \eeq
 Relation (\ref{eq:Etherington}) allows one to calculate both distances by solving the GDE with the initial conditions (\ref{eq:ID2}) imposed either
 at the source or at the observation point. 
 
 In this paper we have found it much simpler to impose the initial conditions
 at the location of the source, and to
 integrate the equations forward in time. Moreover, instead of solving the GDE directly, we simply use the geodesic tracker
and follow directly two additional null geodesics $\gamma_1(\lambda)$ and $\gamma_2(\lambda)$, slightly perturbed with respect to the principal one, which we denote with $\gamma_0(\lambda)$.  
We specify the initial conditions for them at the source:
\bea
x^a_1(\lambda_{\cal S}) &=& x^a_2(\lambda_{\cal S}) = x^a_0(\lambda_{\cal S}) \\
p_1^a(\lambda_{\cal S}) &=& p_0^a(\lambda_{\cal S}) + \epsilon \xi_1^a(\lambda_{\cal S}) \\
p_2^a(\lambda_{\cal S}) &=& p_0^a(\lambda_{\cal S}) + \epsilon \xi_2^a(\lambda_{\cal S})
\eea
where $x^a_I$ are the coordinates of geodesic $\gamma_I$ and $p^a_I$ is its 4-momentum.
We can then compute $\calD\UD{A}{B}$ by using the fact that:
\bea
\calD\UD{A}{B}(\lambda) &=& \lim_{\epsilon \to 0} \frac{\sqrt{g(\lambda_{\cal S})}}{\epsilon} \left [
  \begin{array}{ll}
    g_{ab}(x^a_1-x^a_0)\,\xi_1^b \qquad & g_{ab}(x^a_2-x^a_0) \,\xi_1^b \\[0.5cm]
    g_{ab}(x^a_1-x^a_0)\,\xi_2^b \qquad & g_{ab}(x^a_2-x^a_0) \,\xi_2^b
  \end{array}
\right ]
\eea
where $g(\lambda_{\cal S})$ is the determinant of $g_{ab}$ at the geodesic initial
location.
This is the approach we take in the computations described in Section~\ref{sec:results}.

\subsection{Homogeneous cosmologies}
This formalism takes on a particularly simple form in the exactly
homogeneous and isotropic cosmological models (the FLRW class),
defined by the line element:
\beq
d s^2 = - d t^2 + a(t)^2 dl^2
\eeq
where $dl^2$ is the line element of one of the three three-dimensional
constant-curvature spaces of Euclidean signature.
In this case, geodesics can move along coordinate lines and be parametrized
by the coordinate time. In the flat case,
for instance, we can
choose $x$ as the geodesic direction (so that $\xi_1^a=a(t) \delta_y^a$ and
$\xi_2^a=a(t) \delta_z^a$, where $a(t)$ is the scale factor). The 
matrix $\calD\UD{A}{B}$ is then given by:
\bea
\calD\UD{A}{B}(t) &=& a_{\cal S} \left [
  \begin{array}{cc}
    a(t) x(t) & 0 \\ 
    0         & a(t) x(t)
  \end{array}
\right ]
\eea
where $x(t)$ is the coordinate distance travelled along the geodesic
at time $t$:
\beq
x(t)=\int_{t_{\cal S}}^{t} \frac{dt}{a(t)}
\eeq 
Given the initial normalization $u_{\cal S}^a\,p_a=-a_{\cal S}^{-1}$, equation (\ref{eq:DL}) 
becomes:
\beq
  \label{eq:flrwDL}
  D_{\rm L}= a_{\cal O} (1+z) \int_{t_{\cal S}}^{t_{\cal O}} \frac{dt}{a(t)}
\eeq
Noticing that, in an FLRW model, the redshift 
$z$ only depends on the ratio between the scale factor at the time of 
detection and the scale factor at the time of emission:
\beq
z=\frac{a(t_{\cal O})}{a(t_{\cal S})} - 1,
\eeq
it is easy to show that equation (\ref{eq:flrwDL}) coincides with the
usual textbook expression for $D_{\rm L}$, which we quickly recall.
We first need to calculate the comoving distance covered by a photon between 
${\cal S}$ and ${\cal O}$:
\beq
D_{\rm M}(z)=a_{\cal O} \int_{t_{\cal S}}^{t_{\cal O}} \frac{dt}{a(t)} = (1+z) S\left(\Omega_k,\int_0^z \frac{d\zeta}{H(\zeta)(1+\zeta)^2}\right),
\eeq
with 
\beq
H(\zeta) = H_{\cal S}\sqrt{\Omega^{\cal S}_{\rm M}(1+\zeta)^{-3}+\Omega^{\cal S}_\Lambda+\Omega^{\cal S}_k(1+\zeta)^{-2}},
\eeq
and
\bea
S(k,x) &=& \left\{
  \begin{array}{ll}
  k^{-1/2} \sin k^{1/2} x  & \textrm{for } k > 0 \\
              x  & \textrm{for } k = 0 \\
  |k|^{-1/2} \sinh |k|^{1/2} x & \textrm{for } k < 0
  \end{array}
\right.
\eea
Notice that the reference values for all quantities are those at the source:
$a_{\cal S}$, $H_{\cal S}$, and $\Omega^{\cal S}_{\rm X}$ are the model's scale 
factor, Hubble rate, and density parameters at the time the photon is emitted, 
respectively. As is customary, we also define the curvature $\Omega$ parameter by:
\beq
\Omega^{\cal S}_k=1-\Omega^{\cal S}_{\rm M}-\Omega^{\cal S}_{\Lambda}.
\eeq
Notice that referring to the initial values of these parameters rather
than the final ones changes our expressions from the standard textbook 
treatment. It is straightforward to show that the usual formulae are
recovered if one expresses all quantities at the source in terms of the 
corresponding ones at the observer.

Having found an expression for $D_{\rm M}(z)$, we can use it to derive the apparent 
luminosity $\ell$ of an object of intrinsic luminosity ${\cal L}$
(for details, see e.g.~\cite{Hogg:1999ad}):
\beq
\ell = \frac{\cal L}{4 \pi D_{\rm M}(z)^2 (1+z)^2}.
\eeq
Since the apparent luminosity is defined as:
\beq
D_{\rm L} (z) = \sqrt{\frac{\cal L}{4 \pi \ell}},
\eeq
we finally obtain:
\beq
\label{eq:ldflrw}
D_{\rm L} (z) = D_{\rm M}(z) (1+z) = (1+z)^2 S \left( \Omega_k, \int_0^z \frac{d\zeta}{H(\zeta)(1+\zeta)^2} \right)
\eeq
This can be easily identified, on a flat background, with (\ref{eq:flrwDL}).
In homogeneous and isotropic cosmologies, therefore, the luminosity distance
only depends on the redshift, and is parametrized by global quantities such 
as the matter density and the curvature of spatial slices.
In the Einstein-de Sitter (EdS) model, $D_{\rm L}$ simply reduces to
\beq
\label{eq:ldeds}
D_{\rm L} (z) = \frac{2 (1+z)^2}{H_{\cal S}} \left( (1+z)^{1/2}-1 \right) 
\eeq

\subsection{Inhomogeneous cosmologies}
The propagation of light in lumpy spacetimes has been studied since the
1960's with various approaches, starting with the EBA
proposed in~\cite{1964SvA.....8...13Z} and later
generalized in~\cite{1972ApJ174L115D,Dyer:1973zz}.
The key idea inspiring these studies is that, in cosmological models
where the matter is distributed in lumps, a large fraction of the light
beams would not contain matter, and would therefore not be affected
by the Ricci focusing characteristic of their FLRW counterparts.

Other limitations of the FLRW approximation and the related physical
effects were subsequently analysed, both in approximate scenarios and
in exact cosmological models (typically belonging to the Swiss-Cheese 
family)~\cite{Fleury:2014gha,Seitz:1994xf,
KristianSachs,1967ApJ150737G,1969ApJ...155...89K,
1970ApJ...159..357R,Lamburt2005,2010PhRvD..82j3510B,
2011PhRvD..84d4011S,Nwankwo:2010mx,2012MNRAS.426.1121C,
2012JCAP...05..003B,Lavinto:2013exa,Troxel:2013kua,
Bagheri:2014gwa,
Peel:2014qaa}.
A few robust features of these studies, that do not depend on
the details of the models used, include that:
\bit
\item Light sources appear reduced in size and dimmer in a lumpy 
spacetime than in a homogeneous one with the same mean density;
\item The angular distance does not have a maximum, but keeps growing
all the way to the cosmic horizon;
\item The actual deceleration parameter $q_0$ 
is larger
than in 
the case where the same data is analysed with an FLRW model with the
same mean density.
\eit
Later, when we measure the $D_{\rm L}(z)$ relationship in BHL spacetimes,
we will use these features as guidelines for what to expect. Many of
them do indeed hold for such highly nonlinear spacetimes too.

In fact, as discussed at length in Section~\ref{sec:results}, the luminosity
distance in a BHL follows rather closely the EBA~\cite{1964SvA.....8...13Z},
which we report for completeness:
\beq
\label{eq:eb}
D_{\rm L}(z)=\frac{2(1+z)^2}{5 H_{\cal S}}\left( 1 - \frac{1}{(1+z)^{5/2}}\right).
\eeq
In Section \ref{sec:bhl} we will explain why the EBA
is a good approximation of the redshift--luminosity 
distance in a BHL, and point out that it is equivalent to neglecting the
Ricci term in the standard geodesic deviation equation.

\subsection{Geodesics and observer classes}

As with many other quantities of interest that can be calculated in 
inhomogeneous cosmologies, the calculation of $D_{\rm L}(z)$
requires the choice of a time coordinate. In general, representing 
the spacetime in the geodesic gauge will lead to coordinate observers 
which are diversely affected by neighbouring gravitational structures, 
and may experience, e.g., light redshifting which has nothing to do with
a global, suitably defined expansion rate (an observational cosmologist 
would call these {\it local effects}).

A study of light propagation in inhomogeneous spaces, especially one that is 
targeted at the comparison with the FLRW class, is then left with two
possibilities: a statistical approach in which observers and sources are
distributed stochastically throughout the spacetime, and a single 
$D_{\rm L}(z)$ relationship is obtained by averaging over
their locations and four-momenta; or the construction of one or more
classes of {\it cosmological} observers, based on geometry-inspired 
considerations such as following the geodesics of the average
gravitational field, or geodesics with minimal deviation.
We find the latter approach more likely to yield insight on the 
different gauge choices and related effects, and therefore use
it in the remaining of this paper. Statistical reasoning is,
however, also an important ingredient, as the observational data
is arguably to be modelled through a mix of different observer and source
states of motion. As statistical analyses are a tricky endeavour in
cosmology, we leave this task for future work.

Notice that the second strategy is particularly difficult to deploy
on vacuum spacetimes, as the sources of the gravitational field are 
only perceived through their effect on the metric tensor, and not 
through the presence of matter, so singling out a ``local''
component of the gravitational field will in some cases not even be 
well defined (for a discussion of this point, see 
e.g.~\cite{Marra:2012pj} or~\cite{lrr-2004-9} and references therein).  
We will however exploit the existence of global (albeit discrete) 
symmetries in our BHLs and only turn our attention to geodesics which
are by construction least affected by local effects: these include, for 
instance, the geodesics running along the edges of the fundamental 
periodic cell constituting the lattice.

\section{Light propagation in BHLs}
\label{sec:bhl}

In this section, we build an approximate model for the propagation of light
in a BHL, based on a perturbative expansion in the BHL compactness parameter.
This will serve as a qualitative analysis of the physics of the propagation of light and as a support in the interpretation of the numerical results
presented in section~\ref{sec:results}. Note that an expansion in a similar parameter has already appeared in the context of BHLs \cite{Bruneton:2012ru}, although the details are different.

\subsection{A perturbative expansion in the compactness parameter}

Let $L$ denote the characteristic size of a lattice cell, such as its initial geodesic length,
and let $M$ be a characteristic mass, i.e.~the total mass contained in a cell. 
As in~\cite{Bentivegna:2013jta}, we can introduce the dimensionless parameter
\bea
\mu = \frac{M}{L}
\eea
measuring the lattice compactness. If we additionally introduce the characteristic mass density
$\rho = M L^{-3}$, we can see that
\bea 
\mu = \rho L^2,
\eea
i.e.~it goes down to zero as we decrease the size of
a cell keeping the mass density of the corresponding FLRW model fixed. Note that $\rho$ is related to the curvature
scale of the Friedmann model via $R=\rho^{-1/2}$, so $\mu$ can be reinterpreted as the separation of scales between the
size of an individual lattice cell and the radius of curvature of the FLRW model:
\bea
\mu = \frac{L^2}{R^2}.
\eea

Note that the definition of $\mu$ involves a certain vagueness: we may take for the mass scale $M$ 
the ADM mass of the black hole measured at the other end of the Einstein-Rosen bridge, but also some other related parameter. 
Also the choice of the length scale involves a certain arbitrariness. At the leading order we expect this ambiguity to be irrelevant.

We will now show how $\mu$ can be used to find a perturbative approximation for the metric tensor of the lattice model.
The approximation is different from the standard perturbative approximation on an FLRW background, in the sense that it
does not require the density contrast $\delta$ of the matter perturbation to be small. Obviously the problem of BHLs lies beyond the validity regime of the cosmological perturbation theory, because in a BHL we are 
dealing with $\delta = -1$ everywhere. 

We begin by introducing a coordinate system on a single cell. Let $\gzero$ denote the background FLRW metric and $x^\mu$ be the Riemannian normal coordinate system around any point $P$. The metric takes the form of
\bea
 \gzero_{\mu\nu} = \eta_{\mu\nu} - \frac{1}{3} R_{\mu\alpha\nu\beta}\Big|_P\,x^\alpha\,x^\beta + O(x^3).
\eea
Since $R$ is the curvature scale of the metric, the coefficients in the expansion above are of order $R^0$ (the flat metric), $R^{-2}$
(the Riemann tensor),
$R^{-3}$ (the next term involving $\nabla_\sigma R_{\mu\alpha\nu\beta}$), and so on. The Taylor expansion
in the Riemannian normal coordinates becomes thus the expansion in negative powers of $R$.
We now introduce the rescaled coordinates $\tilde x^\mu = L^{-1}\,x^\mu$ and the
rescaled Riemann tensor at point $P$ in coordinates $x^\mu$. 
\bea 
 r_{\mu\alpha\nu\beta} = R^2\,R_{\mu\alpha\nu\beta}\Big|_P.
\eea 
Both $\tilde x$ and $r_{\mu\alpha\nu\beta}$ are $O(1)$ in the expansion in $R$, at least within a single lattice  cell
around $P$. The metric $\gzero$ can be expressed in the new coordinates. The 
metric tensor components in those coordinates will be denoted by $\tgzero_{\mu\nu}$,
i.e. 
\bea
\gzero = \gzero_{\mu\nu} \,\dd x^\mu\otimes\dd x^\nu =
\tgzero_{\mu\nu}\,\dd \tilde x^\mu\otimes\dd \tilde x^\nu.
\eea
Its expansion in $\tilde x^\mu$ takes the form of
\bea
\tgzero_{\mu\nu} = L^2\left(\eta_{\mu\nu} - \frac{\mu}{3} r_{\mu\alpha\nu\beta}\,\tilde x^\alpha\,\tilde x^\beta + O(\tilde x^3 L^3)\right).
\eea
The first of the remaining higher-order terms $\nabla_\sigma R_{\mu\alpha\nu\beta} \,L^{3}\,\tilde x^\sigma\,
\tilde x^\alpha\,\tilde x^\beta$ contains the covariant derivative $\nabla_\sigma R_{\mu\alpha\nu\beta}\Big|_P$, which is 
$O(R^{-3})$ as we noted before. Therefore the whole term in question can be re-expressed as $r_{\sigma\mu\alpha\nu\beta}\,\tilde x^\sigma\,
\tilde x^\alpha\,\tilde x^\beta\,\frac{L^3}{R^3}$, where we have defined by analogy 
the rescaled derivative of the curvature $r_{\sigma\mu\alpha\nu\beta} = R^{3}\,\nabla_\sigma R_{\mu\alpha\nu\beta}$, which again is $O(1)$ in $R$. We see that the whole term turns out to be $O(\mu^{3/2})$. Similar reasoning can be applied to all higher terms, yielding
higher powers of the dimensionless parameter $\mu$.
We thus see that
\bea
\tgzero_{\mu\nu} = L^2\left(\eta_{\mu\nu} - \frac{\mu}{3} r_{\mu\alpha\nu\beta}\,\tilde x^\alpha\,\tilde x^\beta + O(\mu^{3/2} )\right),
\eea
i.e. in the rescaled coordinates the expansion in the negative powers of $R$ turns in a natural way into an expansion in powers of  $\mu$,
valid in a region of size $L$ around $P$.

We can explain the physical meaning of the expansion above in the following way: if the background metric $\gzero$ has
the curvature scale of $R$, then in an appropriately picked, quasi-Cartesian coordinate system $x^\mu$ it has the Taylor expansion in which 
the terms are of increasing order in $R^{-1}$. If we then pick a domain of size $L$, then the metric in this domain, again in 
appropriate coordinates,
has the form of the flat metric plus perturbations from the curvature and its derivatives. A simple way to obtain a perturbation of this kind is to use the Taylor expansion we mentioned before and rescale the coordinates by $L$, which yields an expansion in
powers of $\mu^{1/2}$.

 Now we can add the perturbation due to the discrete matter content. We assume the full metric to be
\bea
 \tilde g_{\mu\nu} = L^2\left(\eta_{\mu\nu} - \frac{\mu}{3} r_{\mu\alpha\nu\beta}\,\tilde x^\alpha\,\tilde x^\beta + 
 \mu\,h_{\mu\nu}\left(\tilde x^\alpha\right) + O(\mu^{3/2} )\right) \label{eq:fullg}
\eea
with the perturbation $h_{\mu\nu}\left(\tilde x^\alpha\right)$ of order $O(1)$ in $\mu$. Note that 
the dependence on $\tilde x^\mu$ means that the characteristic physical size of the perturbation is the size of a cell, i.e.~$L$.
The Einstein tensor of the metric above is
\bea
G_{\mu\nu}\left[\tilde g_{\alpha\beta}\right] = G_{\mu\nu}\left[\tgzero_{\alpha\beta}\right] +
\mu\,G'_{\mu\nu}\left[h_{\alpha\beta}\right]\left(\tilde x^\alpha\right) + O(\mu^{3/2}),
\eea 
where $G'_{\mu\nu}[\cdot]$ is the linearisation of the Einstein tensor around a flat metric $\eta_{\mu\nu}$. In particular,
in the harmonic gauge it is simply $-\frac{1}{2}\Box h_{\alpha\beta}$. 
We now return to the original, unrescaled
coordinate system, where this equation takes the form of
\bea
G_{\mu\nu}\left[g_{\alpha\beta}\right] = G_{\mu\nu}\left[\gzero_{\alpha\beta}\right]+
\rho\,G'_{\mu\nu}\left[h_{\alpha\beta}\right]\left(x^\alpha / L\right)  + O(\mu^{3/2}),
\eea 
i.e.~the perturbation of the Einstein tensor is $O(\rho)$, just like the Einstein tensor of the FLRW metric.
It means that this approximation works even if the density perturbation is of the order of the background energy density.
We may therefore use $h_{\mu\nu}$ to cancel the stress-energy tensor of the underlying FLRW metric everywhere except
on a single worldline.

Recall that $G_{\mu\nu}\left[\gzero_{\alpha\beta}\right] = 8\pi G \rho \, u_{\mu} u_{\nu}$, where $u^{\mu} = (1,0,0,0)$ is the cosmic fluid 4-velocity.
We impose the linear PDE on the metric perturbation:
\bea
G'_{\mu\nu}\left[h_{\alpha\beta}\right] = 8\pi G \left(-1 + C\delta^{(3)}(x^\alpha)\right) u_{\mu}\,u_{\nu}
\eea 
 with periodic boundary conditions and with the constant $C$ chosen so that the RHS integrates out to zero
 over one cell. The solution can be obtained using Appell's $\zeta$ function~\cite{Steiner:2016tta}.
 It diverges at the centre, where the approximation fails, but near the cell's boundary it is likely to work well.
 The resulting approximate metric is vacuum everywhere and periodic.
 \subsection{The continuum limit}
 Let us now consider the metric (\ref{eq:fullg}) along with its Christoffel symbols and Riemann tensor. It is straightforward
 to see that
 \bea
 \tilde g_{\mu\nu} &=& \tgzero_{\mu\nu} + L^2\,\mu\,h_{\mu\nu}(\tilde x^\rho) \\
 \Gamma\UD{\alpha}{\beta\gamma} \left[\tilde g_{\kappa\lambda}\right]&=&
 \Gamma\UD{\alpha}{\beta\gamma} \left[\tgzero_{\kappa\lambda}\right] +
\mu\,{\Gamma'}\UD{\alpha}{\beta\gamma}\left[h_{\kappa\lambda}\right](\tilde x^\rho) \\
  R\UD{\alpha}{\beta\gamma\delta} \left[\tilde g_{\kappa\lambda}\right] &=&
 R\UD{\alpha}{\beta\gamma\delta} \left[\tgzero_{\kappa\lambda}\right]  + 
 \mu\, {R'}\UD{\alpha}{\beta\gamma\delta} \left[h_{\kappa\lambda}\right](\tilde x^\rho). 
 \eea
 We can now go back to the original, unrescaled coordinates and obtain
 \bea
 g_{\mu\nu} &=& \gzero_{\mu\nu} + \mu\,h_{\mu\nu} \left(x^\rho/L\right) \label{eq:gpert-original}\\
 \Gamma\UD{\alpha}{\beta\gamma} \left[g_{\kappa\lambda}\right]&=&
 \Gamma\UD{\alpha}{\beta\gamma} \left[\gzero_{\kappa\lambda}\right] +
 \mu^{1/2}\,\rho^{1/2}\,{\Gamma'}\UD{\alpha}{\beta\gamma}\left[h_{\kappa\lambda}\right] \left(x^\rho/L\right) \label{eq:gammapert-original}\\
  R\UD{\alpha}{\beta\gamma\delta} \left[g_{\kappa\lambda}\right] &=&
 R\UD{\alpha}{\beta\gamma\delta} \left[\gzero_{\kappa\lambda}\right]  + 
 \rho\, {R'}\UD{\alpha}{\beta\gamma\delta} \left[h_{\kappa\lambda}\right] \left(x^\rho/L\right)\label{eq:Riemannpert-original}
 \eea
 plus higher order terms in $\mu$. Consider now the limit $\mu \to 0$, i.e.~where the size of the perturbations decreases in comparison
 to the curvature scale of the background FLRW model, or the limit where the compactness $M/L$ vanishes. 
 Obviously we see that the metric tensor and the Christoffel symbols converge to the FLRW values in this case, while
 the curvature does not. This is due to the fact that the metric $g_{\mu\nu}$ is that of a vacuum spacetime for all positive $\mu$, 
 while the FLRW one is not. 
 This is a key observation in the study of the optical properties of a BHL, which are determined by the GDE
 and are therefore sensitive to the form of the Riemann tensor.
 
 To illustrate this point, consider first a null geodesic. It follows from equations (\ref{eq:gpert-original})--(\ref{eq:Riemannpert-original}) above that its equation has the form of a perturbed FLRW geodesic
 \bea
 x^\mu(\lambda) = \tilde x^\mu(\lambda) + \mu^{1/2}\,\delta x^\mu(\lambda).
 \eea
 where the tilde denotes the FLRW solution without the inhomogeneities.
 The parallel transport of a frame along the geodesic has a similar expansion in $\mu$:
  \bea
 e\DU{a}{\mu}(\lambda) = \tilde e\DU{a}{\mu}(\lambda)  + \mu^{1/2}\,\delta e\DU{a}{\mu}(\lambda).
 \eea
 We can now rewrite the GDE in the parallel-propagated frame along the geodesic
 \bea
 \frac{\dd^2 X^a}{\dd\lambda^2} &=& \left(R\UD{a}{bcd} \left[\gzero_{\kappa\lambda}\right]  + 
 \rho\, {R'}\UD{a}{bcd} \left[h_{\kappa\lambda}\right]\right) p^b\,p^c + O(\mu^{1/2}).
 \eea
 We see that, already at the leading order $O(1)$ in $\mu$, we must take into account the full physical Riemann tensor instead of the simple FLRW one. In particular, since the BHLs are vacuum spacetimes,
 we need to solve the Ricci-free GDE and possibly take into account the non-vanishing Weyl tensor along the way in order to calculate the angular and luminosity distance.
 Neglecting the Ricci tensor in the GDE is equivalent to the EBA (for a discussion of this point, see e.g.~\cite{Fleury:2014gha}). We may thus expect the redshift--luminosity 
 relations for BHLs in the continuum limit to be close to the EBA.\footnote{We neglect here the finite-beam-size
 effects which would become large when $\mu$ becomes very small: the beam may at some point become
 wide enough to encompass a large number of
 black holes. In this situation the interaction between the beam and the black holes becomes quite complicated as we
 cannot use the GDE approximation any more.}

 At the $O(\mu^{1/2})$ order we may expect additional corrections to $D_{\rm L}$ and $D_{\rm A}$ due to higher-order contributions to the geodesic equation as 
 well as to the 
 GDE equation. Additionally, at this order we need to take into account the impact of the inhomogeneities on the observers in their free fall. In this work, we will not 
 concern ourselves with a quantitative analysis of these effects, but we will signal their appearance to the reader when appropriate.

\section{Results}
\label{sec:results}
In order to compute the relationship between redshift and luminosity
distance on the spacetime of an expanding BHL, we carry out
the numerical integration of the geodesic equation (with null 
tangent), along with the integration of Einstein's equation
required to obtain the metric tensor. The latter operation is
performed by a code generated with the Einstein Toolkit, based on 
the \texttt{Cactus}~\cite{cactus} software framework along with modules
such as \texttt{Carpet}~\cite{carpet,carpetweb}, \texttt{McLachlan}~\cite{mclachlan,kranc},
and \texttt{CT\_MultiLevel}~\cite{Bentivegna:2013xna},
as already presented in~\cite{Bentivegna:2012ei,Bentivegna:2013jta,Korzynski:2015isa}.
The geodesic integrator, on the other hand, is a new \texttt{Cactus} 
module that we have written.  It implements a 3+1 decomposition of the geodesic
equation in the form given in \cite{Hughes:1994ea} and we have verified it
against several exact solutions, as reported in Appendix~\ref{app:geo}.

\subsection{Initial data and evolution}
As in~\cite{Yoo:2012jz,Bentivegna:2013jta}
we first construct an initial-data configuration by solving the 
Hamiltonian and momentum constraints on the cube $[-L/2,L/2]^3$ with 
periodic boundary conditions. In particular, we choose free data 
corresponding to conformal flatness:
\beq
\gamma_{ij} = \psi^4 \delta_{ij}
\eeq
and set the trace of the extrinsic curvature to
zero around the origin and to a negative constant $K_c$ near the boundaries,
with a transition region starting at a distance $l$ from the origin:
\bea
K_{ij} &=& \frac{1}{3} K_{\rm c} T(r) \gamma_{ij} + \psi^{-2} \tilde A_{ij}\\
T(r) &=& \left\{
  \begin{array}{ll}
  0 & \textrm{for } 0 \le r \le l \\
  \left(\frac{(r-l-\sigma)^6}{\sigma^6}-1\right)^6&\textrm{for } l \le r \le l + \sigma \\
  1 & \textrm{for } l + \sigma \le r
  \end{array}
\right.
\eea
where we choose $l=0.05 L$ and $\sigma=0.4 L$.
We represent the traceless part of the extrinsic curvature as:
\beq
\tilde A_{ij} = \tilde D_i X_j + \tilde D_j X_i - \frac{2}{3} \tilde \gamma_{ij} \tilde D_k X^k
\eeq
and the conformal factor as:
\beq
\psi = \psi_{\rm r} + \frac{M}{2r} (1-T(r)),
\eeq
where $M$ is the bare mass of the central black hole, and solve the constraints for $\psi_{\rm r}$ and $X^i$.
For our basic configuration, we use $L=10$ and $M=1$ as in~\cite{Bentivegna:2013jta}.

We then proceed to the time evolution of $\gamma_{ij}$ and $K_{ij}$ using a 
variant of the BSSN formulation, implemented in the \texttt{McLachlan} module,
and to the concurrent integration of the geodesic equation (\ref{eq:GE}).

\subsection{Computation of geodesics}

In order to compute geodesics in a 3+1 numerical spacetime, we first
perform a 3+1 decomposition of the geodesic equation (\ref{eq:GE}),
\beq
{ \nabla_p} p^a = 0.
\eeq

We decompose the geodesic tangent vector $p^a$ into its components along and orthogonal to the
unit hypersurface normal $n^a$, which we call $\sigma$ and $q^a$, respectively: $p^a = \sigma n^a + q^a$.  The vector $q^a$ is spatial, i.e.~$q^an_a=0$, and $\sigma = -n_a p^a$.  We use an
affine parametrisation, and $p^a$ is normalized as
\begin{eqnarray}
p^a p_a = \kappa,
\end{eqnarray}
with $\kappa = 0$ for null geodesics.
The spatial coordinates, covariant components of the tangent vector and affine parameter of the geodesic, ($x^i$, $q_i$, $\lambda$) satisfy
\begin{eqnarray}
\frac{dx^i}{dt} &=& -\beta^i + (p^0)^{-1} \gamma^{ik} q_k, \label{eqn:geo3+1x} \\
\frac{dq_i}{dt} &=& -p^0 \alpha \alpha_{,i} + q^j \beta^k_{,i} \gamma_{kj} - \frac{1}{2} (p^0)^{-1} q_l q_m \gamma^{lm}_{,i}, \label{eqn:geo3+1q} \\
\frac{d\lambda}{dt} &=& (p^0)^{-1} \label{eqn:geo3+1lambda}
\end{eqnarray}
where
\begin{eqnarray}
p^0 &=& \frac{(q_k q_j \gamma^{kj} - \kappa)^{1/2}}{\alpha}
\end{eqnarray}
is the time component of $p$ in the foliation-adapted coordinate
basis. Note that the derivative is with respect to coordinate time $t$, not the affine parameter $\lambda$.  These equations are the same as those given in \cite{Hughes:1994ea}, and a derivation is outlined in Appendix~\ref{app:geo3+1}.

Given $(x^i, q_i, \lambda)$ at a time $t$,
equations (\ref{eqn:geo3+1x})--(\ref{eqn:geo3+1lambda}) determine their
evolution along a single geodesic. 
The right hand sides of
eqs. (\ref{eqn:geo3+1x})--(\ref{eqn:geo3+1lambda}) are computed by
interpolating the metric quantities $\beta^i$, $\gamma_{ij}$, $\alpha$
from the evolution grid to the point $x^i(t)$ using fourth-order Lagrange
interpolation, and $(x^i(t), q_i(t), \lambda(t))$ is integrated using
a fourth-order Runge-Kutta method using the Cactus \code{MoL} component.
Additionally, the metric and various other quantities of interest are
interpolated to $x^i$, and all quantities are output as curves
parametrised by $t$ for use in any subsequent analysis once the
simulation is complete.

We implement the above prescription in two new Cactus components
\code{Geodesic} and \code{ParticleUtils}.  The former contains the
equations themselves, and the latter provides library-type
functionality for integrating systems of equations along curves.
A few validation tests are provided in Appendix~\ref{app:geo}. 

We now face the crucial task of selecting which geodesics to track.
Let us notice that, on a space filled with periodic cells, symmetry 
reasons imply that an obvious class of cosmological observers is that formed 
by observers sitting at the cell vertices. Due to the symmetry, these observers 
do not exhibit any proper motions on top of the cosmic expansion, and the ratio of the 
proper distances between arbitrary pairs of observers is constant at all times.
For this study, we construct and analyse two geodesics from this class 
(which we will denote $A$ and $B$), starting at 
the vertex $(-L/2,-L/2,-L/2)$, with initial tangents equal to $p_a^A=(p_0^A,1,0,0)$ and 
$p_a^B=(p_0^B,1/\sqrt{2},1/\sqrt{2},0)$ respectively. $p_0^A = - \alpha \sqrt{\gamma^{xx}}|_A$ 
and $p^0_B = - \alpha \sqrt{(\gamma^{xx}+\gamma^{yy}+2\gamma^{xy})/2}|_B$ are
chosen by the geodesic integrator to ensure that the geodesics
are null. The two geodesics are plotted in Figure~\ref{fig:z}.

In order to measure the luminosity distance along geodesics $A$ and $B$, we
evolve two further pairs of geodesics, with spatial directions given by:
\bea
(1,\epsilon,0) \\
(1,0,\epsilon)
\eea
and
\bea
\left(\frac{1-\epsilon}{\sqrt{2}},\frac{1+\epsilon}{\sqrt{2}},0 \right) \\
\left(\frac{1}{\sqrt{2}},\frac{1}{\sqrt{2}},\epsilon\right)
\eea 
with $\epsilon=10^{-3}$, representative of two narrow beams close to each 
original geodesic. We can then construct the redshift and luminosity distance
along the two beams.
Again, we emphasize that, since we keep the source parameters fixed
and observe the time evolution of each geodesic, this setup is different
(but essentially equivalent) to the one usually adopted in cosmology, where
the observer is fixed and sources with different parameters are considered.

As in~\cite{Bentivegna:2013jta},
we run this configuration on a uniform grid with three different resolutions
(corresponding to $160$, $256$, and $320$ points per side) in order to estimate
the numerical error.
All results presented below are convergent to first order, consistently with the
convergence order reported for the geometric variables in~\cite{Bentivegna:2013jta}.
All curves represent the Richardson extrapolation, at this order,
of the numerical data.
The corresponding truncation error (when visible) is indicated by a shaded region
around each curve.

\subsection{Small-redshift behaviour}
For small distances $d$ from the source, we expect the photon redshift and
luminosity distance to behave respectively like 
\bea
z(d) &\sim& H_{\cal S} d \\
D_{\rm L}(d) &\sim& d
\eea
where $H_{\cal S}$ is related to
the first time derivative of the local volume element at the source location:
\bea
\label{eq:Hi}
H_{\cal S} = \left . \frac{{\rm tr}(K_{ij})}{3} \right |_{\cal S}
\eea
(see \cite{KristianSachs}). Figure~\ref{fig:z} shows that this expectation 
is confirmed by our computation.
For large $d$, however, both quantities grow larger than the linear order.
Furthermore, the redshift clearly exhibits a non-monotonic behaviour engendered
by the inhomogeneous gravitational field. This is easy to explain as a small, periodic redshift 
due to the photons climbing a potential hill near the vertices (away from the nearest black holes) and falling
into wells near the edge or diagonal midpoints (closer to the black holes).
Naturally, the two geodesics are 
affected in different ways as they trace different paths through the gravitational
field. 

\bfi[!h]
\bce
\centering
\begin{minipage}[b]{0.6\linewidth}
\includegraphics[width=1.0\textwidth]{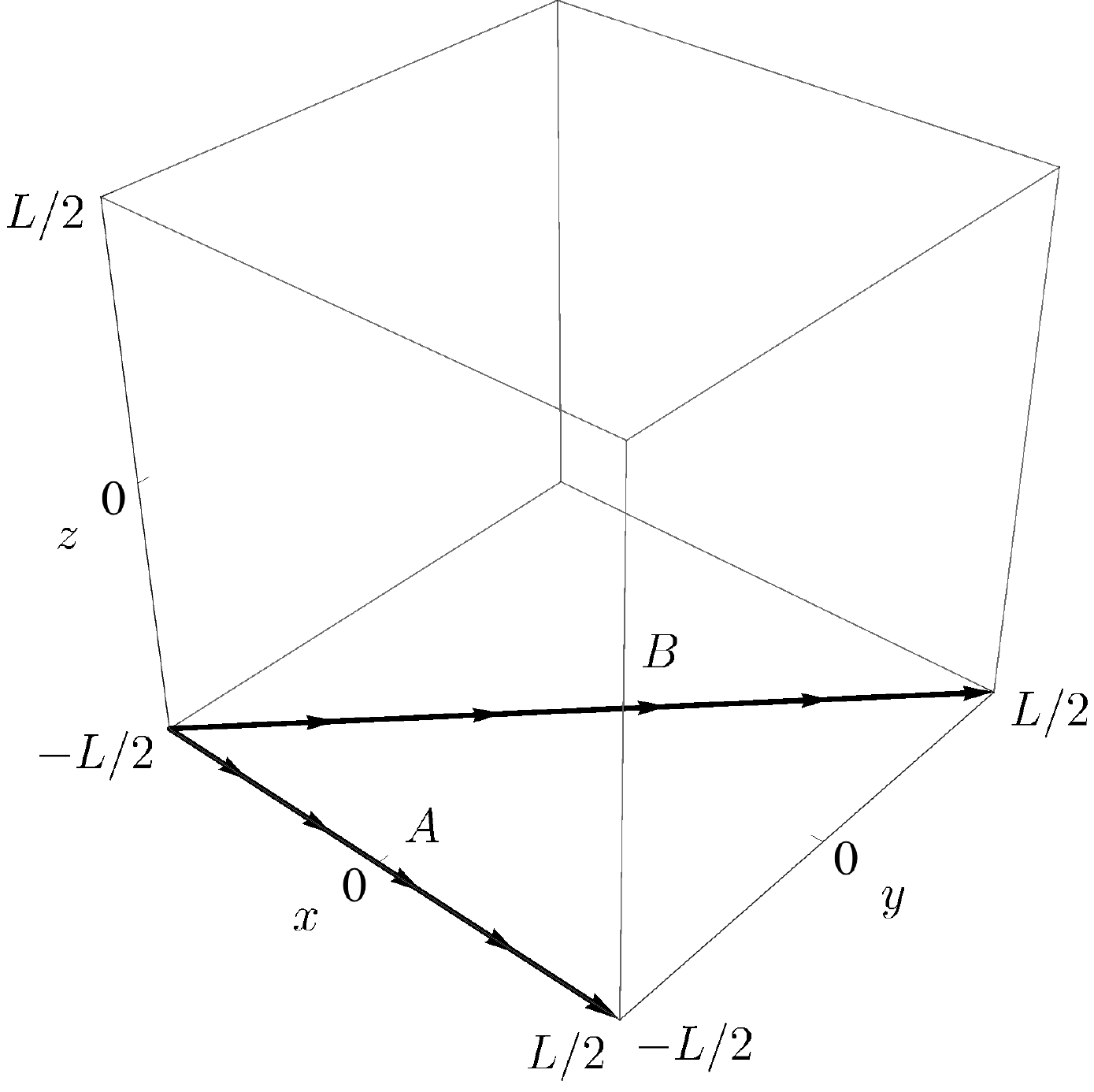}
\includegraphics[width=1.2\textwidth]{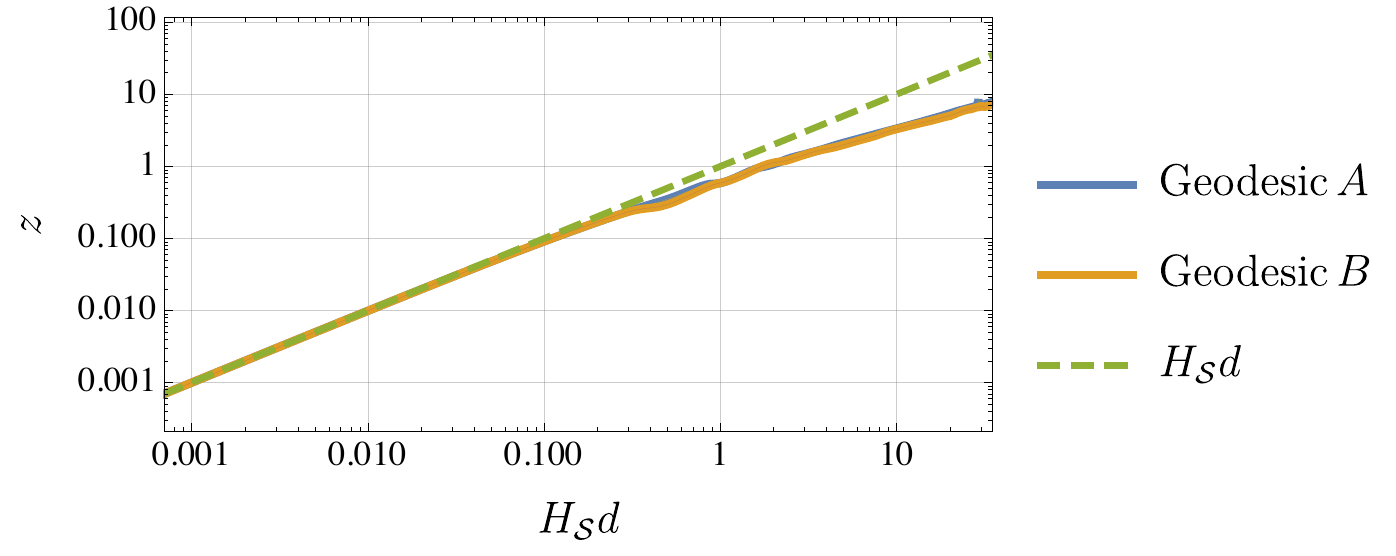}
\includegraphics[width=1.2\textwidth]{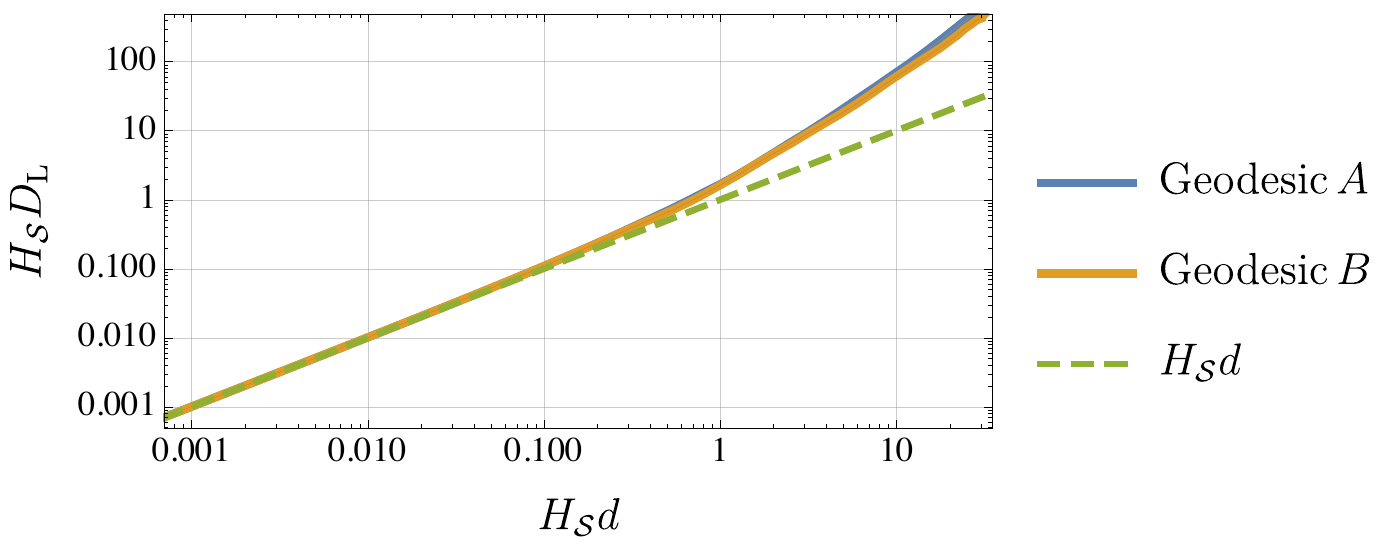}
\end{minipage}
\caption{Top: the paths of geodesics $A$ and $B$ in one of the BHL cells. The 
geodesics run close to the cell edge and diagonal, respectively, at all times. Middle: photon redshift 
as a function of the coordinate distance from the source. Bottom: luminosity
distance as a function of the coordinate distance from the source.
The error bars are indicated by shaded regions (when not visible, they are
included in the width of the curves).
\label{fig:z}}
\ece
\efi

\subsection{Luminosity distance}
Due to numerical error, 
the geodesics deviate from the cell edge and face
diagonal during the evolution, but remain quite close to them (the coordinate separation is less than $0.01\%$ after 
three cell crossings, in both cases).
We can compare the $D_{\rm L}(z)$ relationship for geodesics $A$
and $B$ to the same quantity calculated according to four reference models:
\begin{enumerate}
\item The EdS model (equation (\ref{eq:ldeds}));
\item An FLRW model (equation (\ref{eq:ldflrw}))
with $\Omega_M=0.3$ and $\Omega_\Lambda=0.7$ (henceforth denoted $\Lambda$CDM); 
\item The Milne model~\cite{MILNE01011934}, where redshift and luminosity distance are related by:
\beq
D_{\rm L}(z)=\frac{1}{H_{\cal S}} \frac{z}{(1+z)^2}\left(1+\frac{z}{2}\right);
\eeq
\item The estimate of $D_{\rm}(z)$ via the EBA,
equation (\ref{eq:eb}).
\end{enumerate}

All models are fitted according to two prescriptions:
the initial scale factor $a_{\cal S}$ is always set according to
\beq
a_{\cal S} = {\rm det}(\gamma_{ij})^{1/6}|_{\cal S},
\eeq
while
the initial expansion rate $H_{\cal S}$ is set to either 
(i) 
the initial time derivative
of the proper length of the domain edge (say, the one between $(-L/2,0,0)$ and
$(L/2,0,0$)), which we call a {\it global fit}, and is the same procedure as~\cite{Bentivegna:2013jta};
or (ii) 
equation (\ref{eq:Hi}) (which we call a {\it local fit}). 

\bfi[!h]
\bce
\centering
\includegraphics[width=0.7\textwidth]{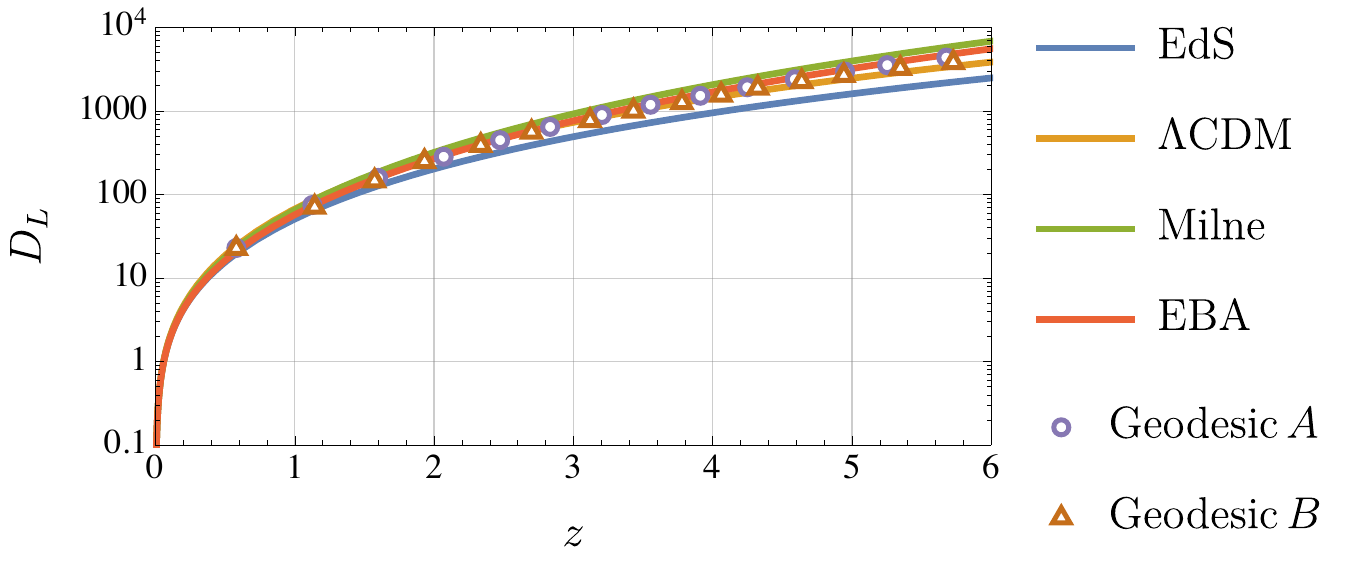}
\includegraphics[width=0.5\textwidth]{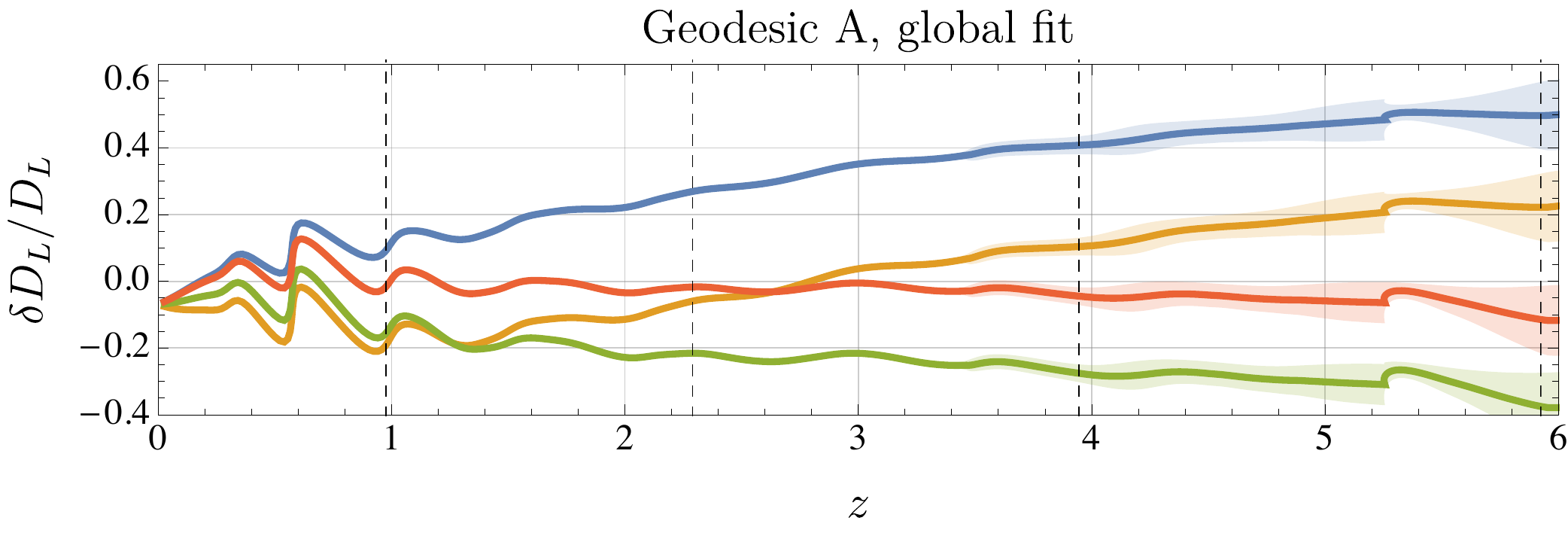} \qquad \qquad \qquad \qquad \\
\includegraphics[width=0.5\textwidth]{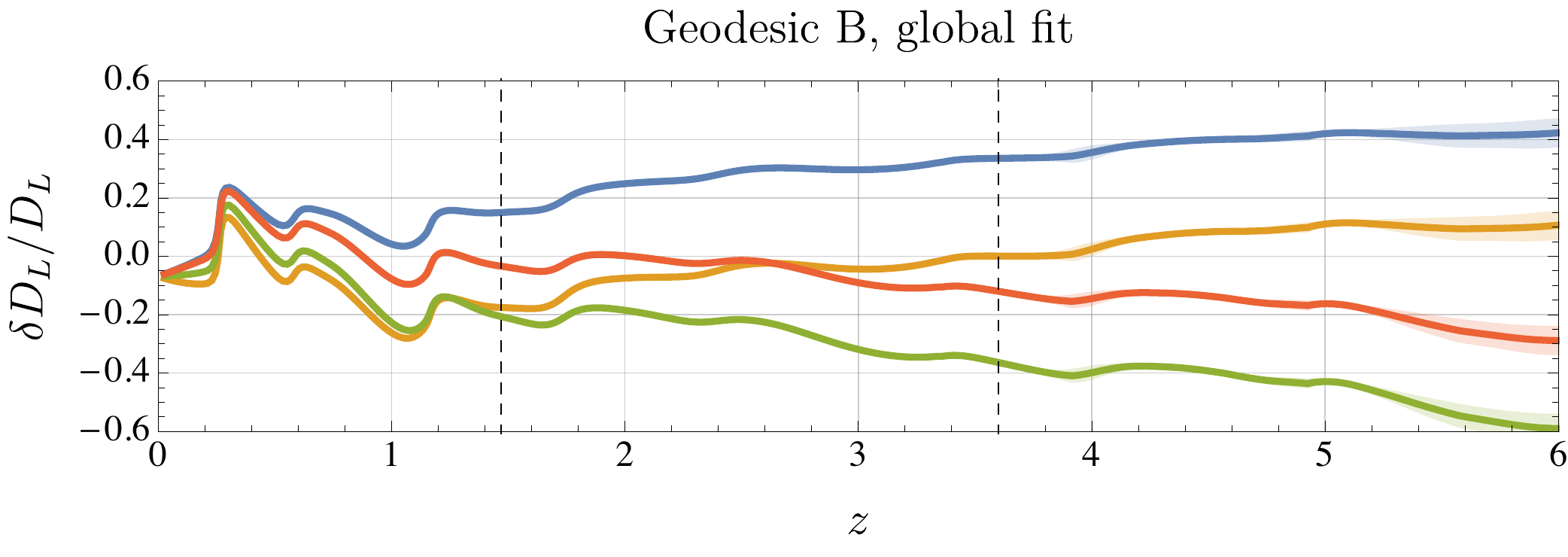} \qquad \qquad \qquad \qquad \\
\includegraphics[width=0.5\textwidth]{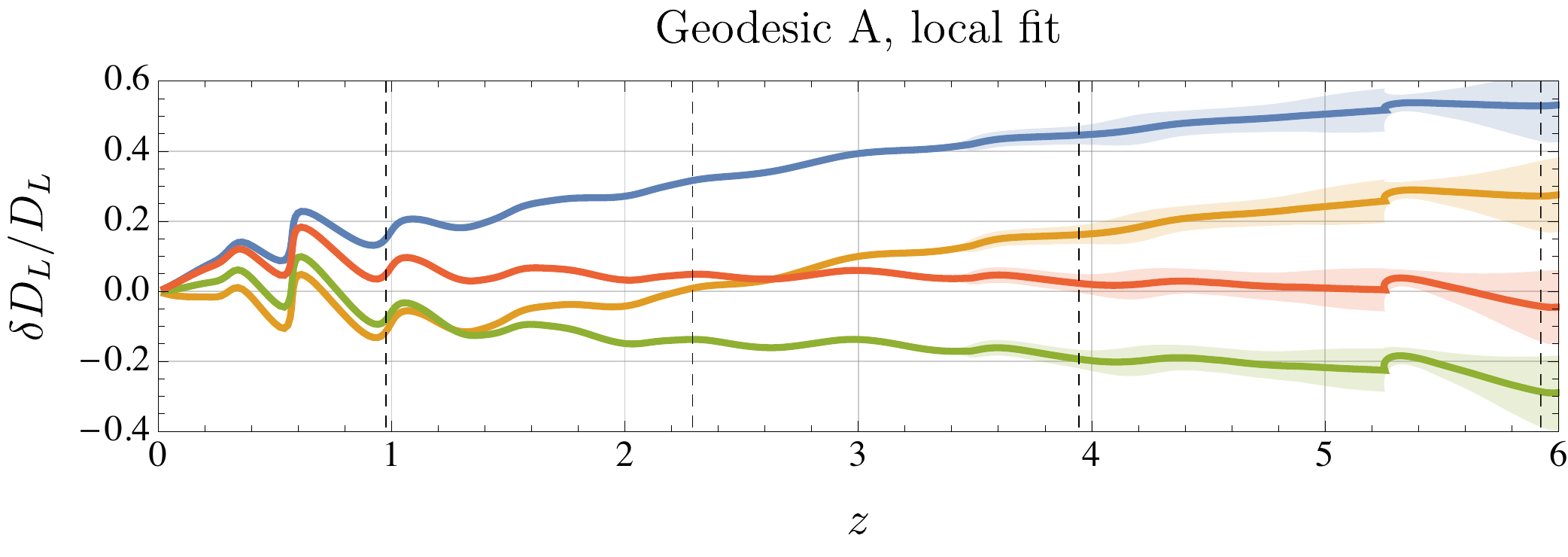} \qquad \qquad \qquad \qquad \\
\includegraphics[width=0.5\textwidth]{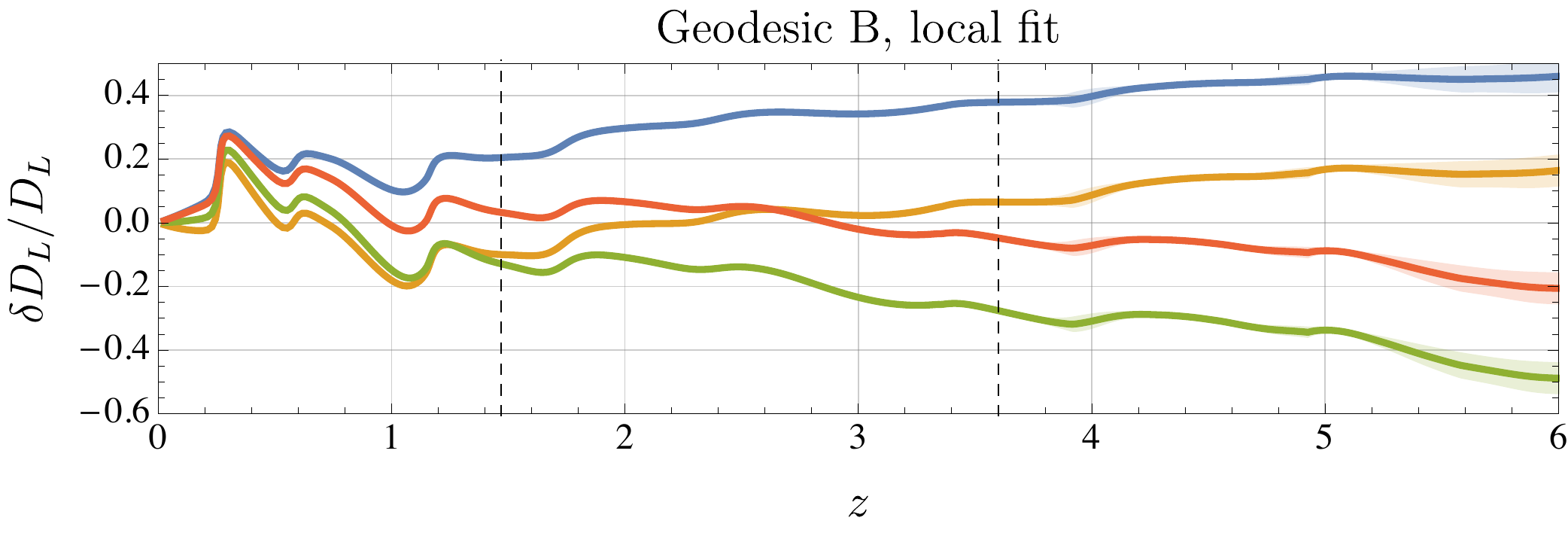} \qquad \qquad \qquad \qquad \\
\caption{Luminosity distance as a function of redshift for geodesics $A$ and $B$ (top plot).
The same relationships in the EdS model, in the $\Lambda$CDM (i.e., FLRW with $\Omega_\Lambda=0.7$
and $\Omega_M=0.3$) model, in the Milne model and in the 
EBA are also plotted. The four models are fitted according to the procedure 
described in~\cite{Bentivegna:2013jta}, using the global expansion rate
computed from the first time derivative of the edge proper length. The relative difference
between the four models and the BHL $D_{\rm L}$ is plotted in the second and third panel.
The fourth and fifth panel illustrate the result of the same procedure, where the 
four models have been fitted using the local expansion rate (\ref{eq:Hi}) instead.
On all plots, the dashed vertical lines mark the points where the geodesics
cross over the periodic boundary.
The error bars are indicated by shaded regions (when not visible, they are
included in the width of the curves or of the data points).
\label{fig:DL}}
\ece
\efi

Figure~\ref{fig:DL} shows all the resulting curves.
We first recall that the expansion of the BHL, measured by the proper
distance of one of its cell edges, could be fitted quite well by an
EdS model with the same initial expasion, as shown in~\cite{Bentivegna:2013jta}.
The two models, however, exhibits markedly
different optical properties. For geodesic $A$, the relative difference 
reaches $60\%$ by redshift $z=6$. This is not surprising: the conditions 
under which these light rays
propagate in a BHL and in an EdS model are substantially different.
In the former case, for instance, null geodesics infinitesimally 
close to $A$ or $B$ accelerate away from, rather than towards, them.

We notice that the EBA provides the best estimate
for $D_{\rm L}(z)$ in a BHL. We conjecture that this result is due to the 
fact that this approximation can capture both the large-scale geometrical
properties of a non-empty universe and the small-scale behaviour of light
rays in vacuum. None of the other models satisfies both these conditions. 
Note also that, for longer times, the EBA works better for the geodesic $A$ 
(along the edge) than for geodesic $B$ (along the face diagonal). This is
easy to explain if we notice that, because of the 4-fold discrete rotational 
symmetry around the edge, there are no
Weyl focusing effects on $A$ and therefore the GDE with the Ricci tensor 
neglected and no Weyl contribution is likely
to be a good approximation for the propagation of the neighbouring light rays. On the other hand along the face diagonal we may expect 
a non-vanishing Weyl lensing around the midpoint area due to the tidal distortion of the rays. Such an effect is not taken into account in the EBA.

\subsection{Fitting the FLRW class}
It is tempting to consider an FLRW cosmology with the same matter content and 
initial expansion as the reference EdS, plus an additional stress-energy 
contribution coming from a cosmological constant, and attempt to tune its value
to reproduce the luminosity distance in the BHL. 

The left panel of
Figure~\ref{fig:OLfit} shows a plot of the required $\Omega_\Lambda$ at each $z$,
for values of $\Omega_M$ in $[0.2,1]$. The right panel shows a cross section of
this surface with the planes $\Omega_M=1$ and $\Omega_M^{\rm eff}=8 \pi/(3 H_{\cal S}^2 L_{\rm prop}^3)$,
where $L_{\rm prop}$ is the initial proper length of a cell edge.
Notice, however, that none of these models would reproduce the expansion history of 
the BHL spacetime, which follows closely that of a region of an EdS
model ($\Omega_M=1$ and $\Omega_\Lambda=0$) with the same $L_{\rm prop}$ and $H_{\cal S}$,
as discussed in~\cite{Bentivegna:2013jta}. This is the core of the fitting problem: the mapping between
different properties of an inhomogeneous spacetime to the FLRW class will be
different, and in general it will not be possible to identify a single
FLRW counterpart capable of reproducing all of the dynamical and optical aspects
of an inhomogeneous cosmology.

\bfi
\bce
\includegraphics[width=0.45\textwidth]{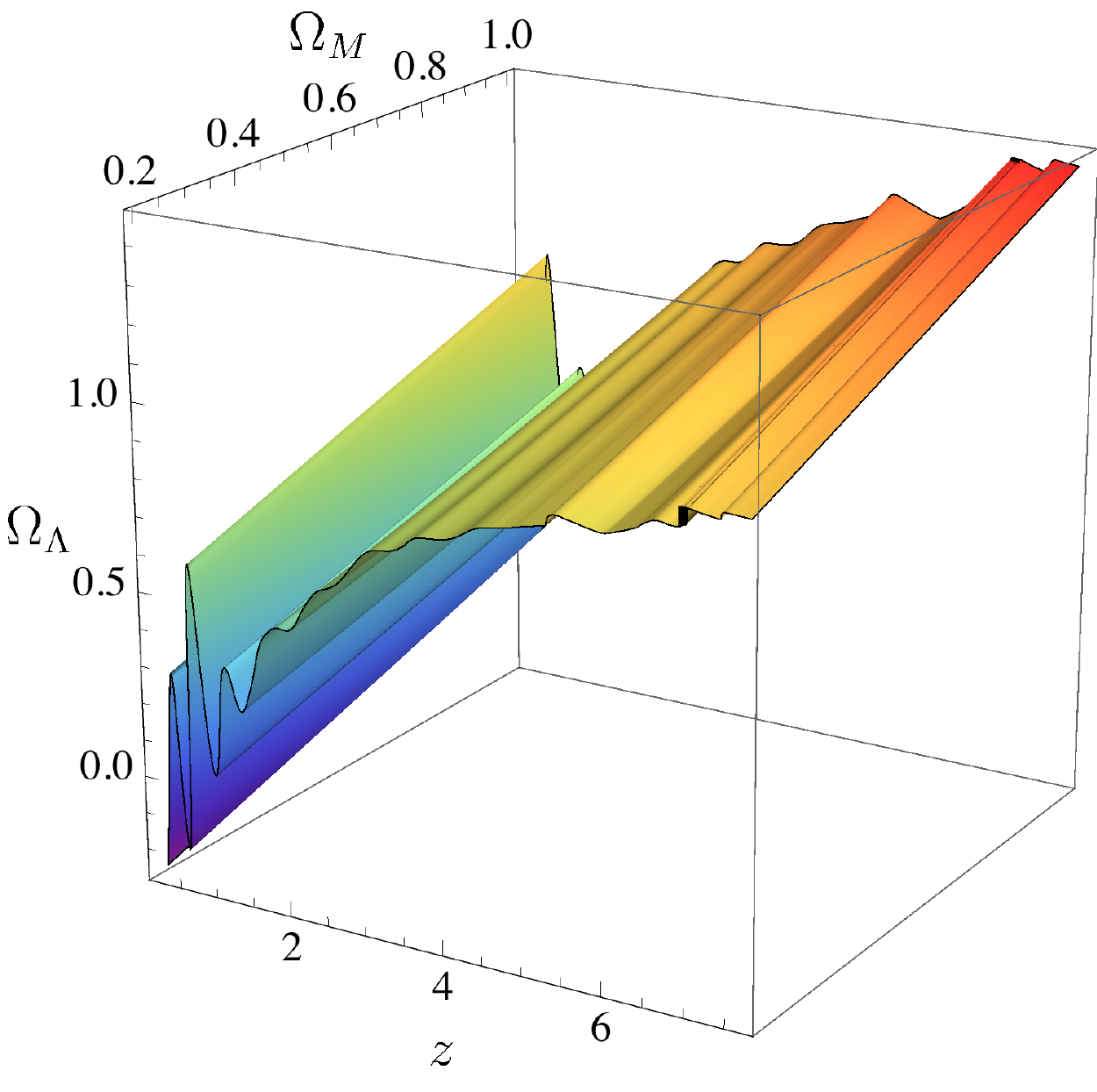}
\includegraphics[width=0.45\textwidth]{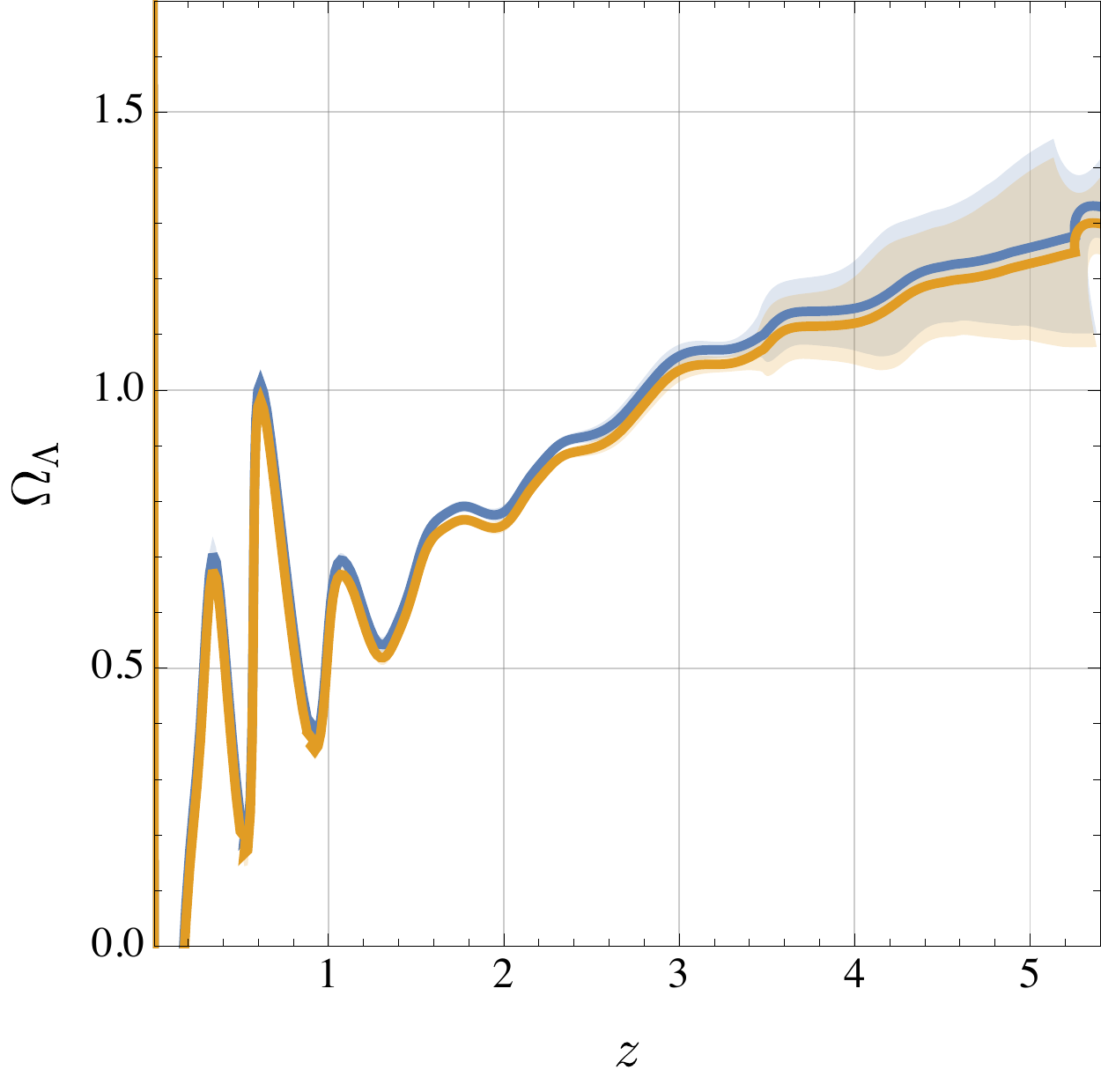}
\caption{Value of $\Omega_\Lambda$ in the best-fit FLRW cosmology, based on the 
luminosity distance measured on geodesic $A$ (left), and its cross sections
with the planes $\Omega_M=1$ and $\Omega_M=\Omega_M^{\rm eff}=8 \pi/(3 H_{\cal S}^2 L_{\rm prop}^3)$
(curve yellow and blue, respectively, on the right plot).
The error bars are indicated by shaded regions (when not visible, they are
included in the width of the curves).
\label{fig:OLfit}}
\ece
\efi

In Figure~\ref{fig:OLfitglob}, we show the constant-$\Omega_\Lambda$ models 
which best fit the $D_{\rm L}(z)$ curves for geodesics $A$ and $B$.
They are obtained for $\Omega_\Lambda^A=1.225$ and $\Omega_\Lambda^B=1.103$,
respectively. The relative difference between these models and the exact solution
is largest around $z=1$, where it reaches $30\%$.

\bfi
\bce
\includegraphics[width=0.7\textwidth]{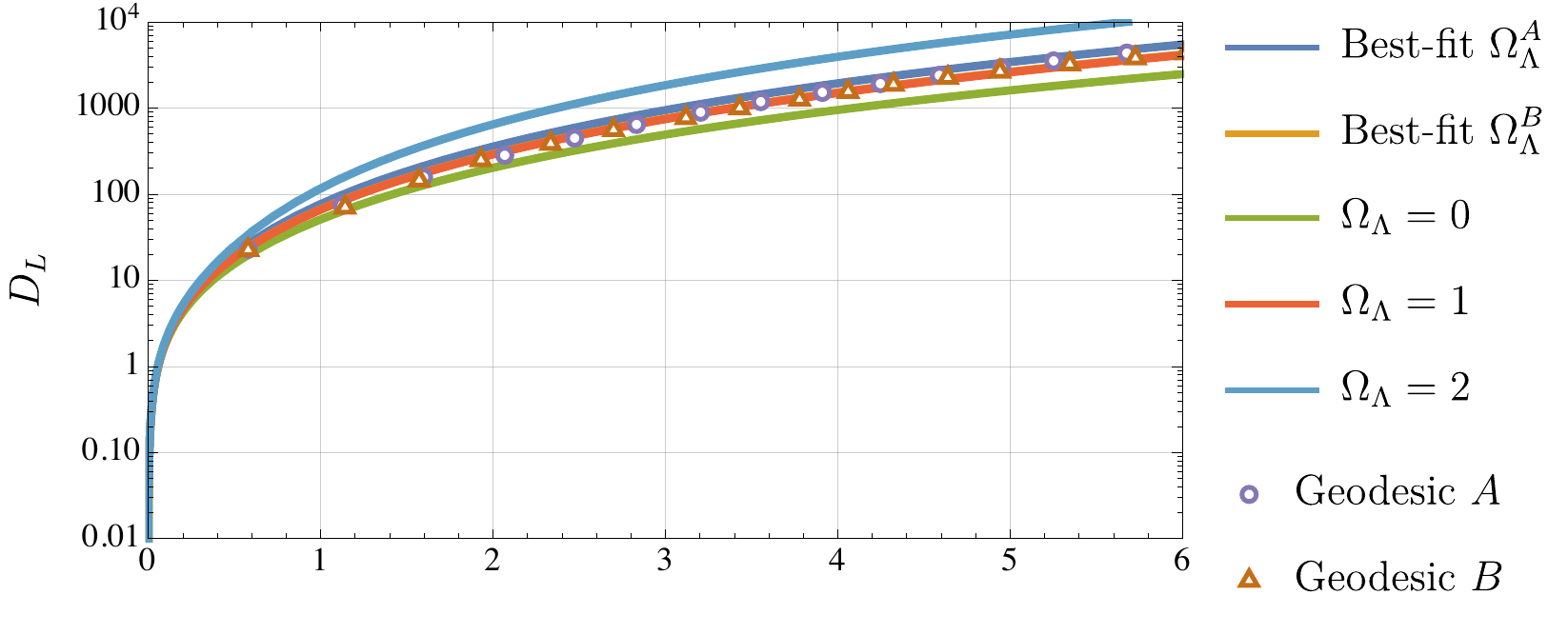}
\includegraphics[width=0.5\textwidth]{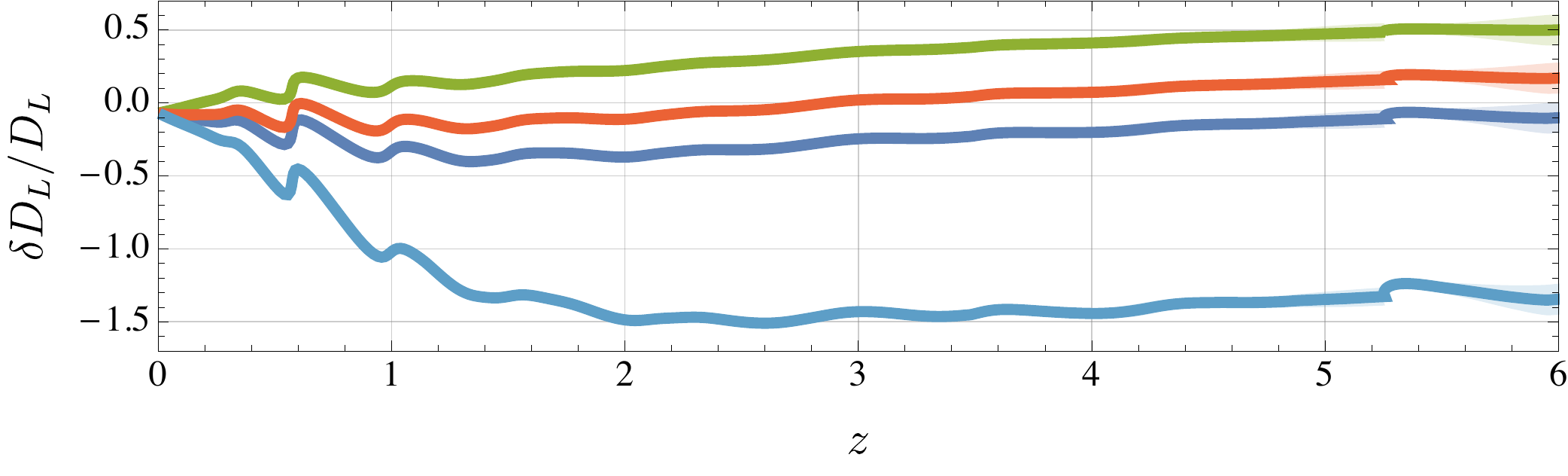} \qquad \qquad \qquad \qquad \qquad \\
\includegraphics[width=0.5\textwidth]{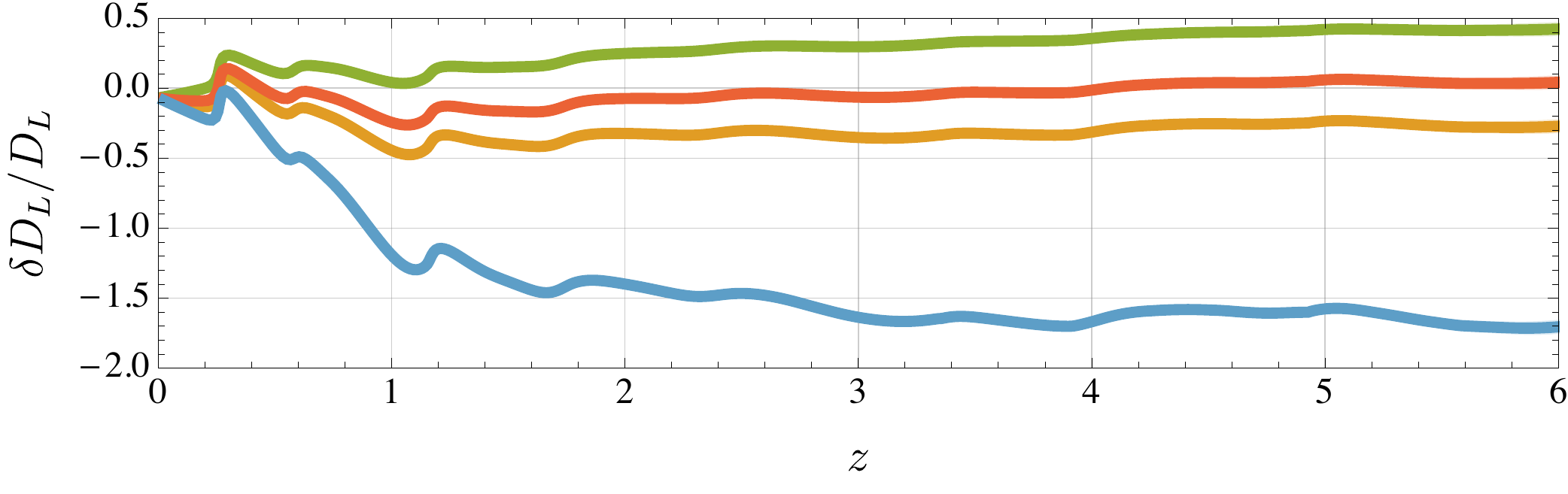} \qquad \qquad \qquad \qquad \qquad \\
\caption{$D_{\rm L}(z)$ for an FLRW model with $\Omega_{\rm M}=1$, 
and $\Omega_\Lambda$ equal to the best-fit values $\Omega_\Lambda^A=1.225$ 
and $\Omega_\Lambda^B=1.103$, as well as to a few other representative
values.
The best-fit models differ from the BHL $D_{\rm L}(z)$ at the $20\%$ level.
The error bars are indicated by shaded regions (when not visible, they are
included in the width of the curves or of the data points).
\label{fig:OLfitglob}}
\ece
\efi

Notice that essentially all quantities discussed so far are affected by oscillations
with a substantial initial amplitude, which is subsequently damped. Similarly
to the oscillations in the redshift, we conjecture that these features are due
to the inhomogeneous gravitational field, and in particular to radiative
modes which likely originate in the oversimplified
initial-data setup we employed. In a space without an asymptotically-flat
region, it is of course difficult to test (or even formulate) this conjecture rigorously. The
compactness of the spatial hypersurfaces, furthermore, means that one cannot
simply ignore this initial transient as is customary in, e.g., binary-black-hole
simulations, as the waves cannot escape from the domain (although their 
amplitude is significantly attenuated by the expansion).
The presence of this unphysical component of the gravitational field, which we 
could barely notice in the length scaling we measured~\cite{Bentivegna:2013jta},
affects very prominently, on the other hand, the BHL optical properties,
and in particular the photon redshift. Better initial-data constructions
which are free from these modes are an interesting field of investigation
which goes beyond the purpose of this work.

Finally, it is worth observing that, as mentioned in Section~\ref{sec:lprop},
different observers would measure a different luminosity distance on the same
spacetime, thereby potentially bringing the BHL result closer to the EdS
curve. A boost with respect to the lattice would, for instance, lower the value
of the distance, according to equation (\ref{eq:DA}). So would a stronger
gravitational field, as would be the case if an observer was located closer
to the centre of a lattice cell.

\subsection{Continuum limit $\mu \to 0$}
Finally, it is instructive to study how this behaviour depends on how tightly packed the BHL
is, as represented by the quantity $\mu=M/L$ introduced in Section~\ref{sec:bhl}.
For simplicity, here we use the bare mass of the central black hole as an estimate of
$M$, and the coordinate size of a cell edge as $L$.
In order to keep $M/L^3$ constant at the value of our base configuration (which had
$M=1$ and $L=10$), we need to have $\mu=M^{2/3}/10$. 
As representative masses we choose $M=\{1/100,1/8,1/2,1,5\}$;
various properties of this BH family are illustrated in Table~\ref{tab:cont}.

\bce
\bt[!b]
\centering
\caption{The bare mass $M$, coordinate size of a cell edge $L=10 M^{1/3}$, its proper size $L_{\rm prop}$,
and the compactness parameter $\mu=M^{2/3}/10$ for a constant-density family of BHLs.
\label{tab:cont}}
\btb{|c|c|c|c|}
\hline
          $M$   &    $L$                  & $L_{\rm prop}$          & $\mu$                  \\
\hline
0.010           &   \phantom{0}2.15       &    \phantom{0}2.73      &   0.0046               \\
0.125           &   \phantom{0}5.00       &    \phantom{0}6.28      &   0.0250               \\
0.500           &   \phantom{0}7.94       &    \phantom{0}9.84      &   0.0630               \\
1.000           &   10.00                 &    12.26                &   0.1000               \\
5.000           &   17.10                 &    21.77                &   0.2924               \\ 
\hline
\etb
\et
\ece

We plot the luminosity distance as a function of $\mu$ in Figures~\ref{fig:mu}
and~\ref{fig:murow}. We observe, in particular, that the difference between the
luminosity distance in a BHL and in an appropriately fitted EdS does not tend to 
zero as $\mu \to 0$. The EdS model, therefore, can reproduce the large-scale 
expansion history of a BHL (as illustrated numerically in~\cite{Yoo:2013yea,
Bentivegna:2013jta}, and deduced analytically in~\cite{Korzynski:2013tea}), 
but is unable to fit its optical properties, even in the limit $\mu \to 0$.

The numerical result is in agreement with the result of the perturbative analysis of Section~\ref{sec:bhl},
where we identified $O(1)$ differences in the GDE of a BHL with respect to 
that of an FLRW model.
This indicates that cosmological-distance estimates of a lumpy spacetime
based on a fit with the FLRW class will exhibit a systematic error, 
\emph{regardless of how lumpy the spacetime is}.
These effects are substantially, but not exhaustively, captured by the EBA, as 
already observed in the case of other inhomogeneous spacetimes~\cite{2012MNRAS.426.1121C,2012JCAP...05..003B,Fleury:2014gha}. 

\bfi
\bce
\includegraphics[width=0.45\textwidth]{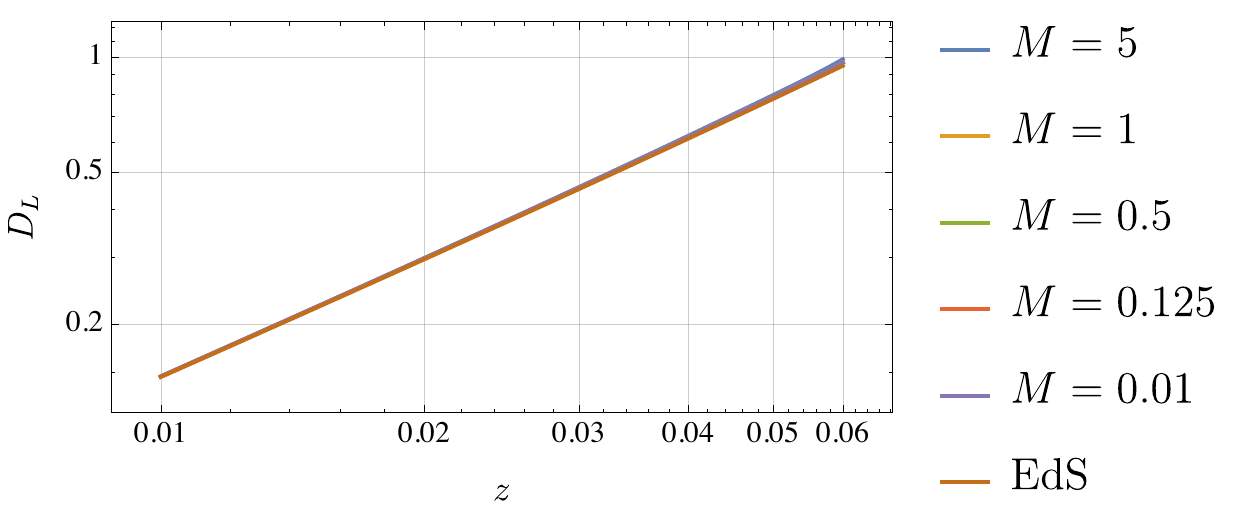}
\includegraphics[width=0.45\textwidth]{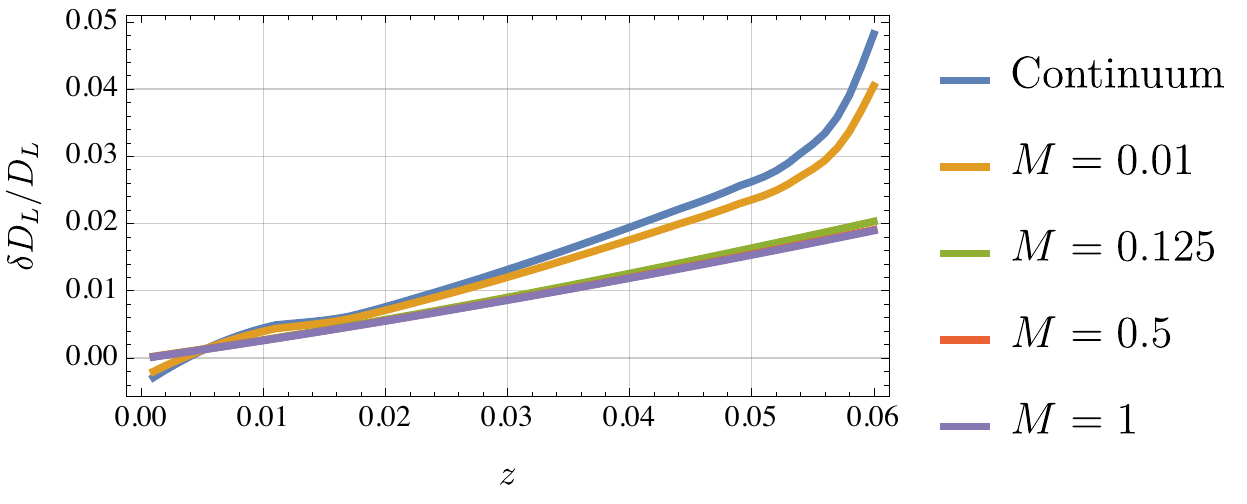}
\caption{Left: luminosity distance for a family of BHLs with the same density
but varying $\mu$.
Right: residual with respect to the EdS model (fitted via the local
expansion rate) of the four lowest-mass models along with their extrapolation
for $\mu \to 0$.
\label{fig:mu}}
\ece
\efi
\bfi
\includegraphics[width=1.0\textwidth, trim=60 0 60 0, clip=true]{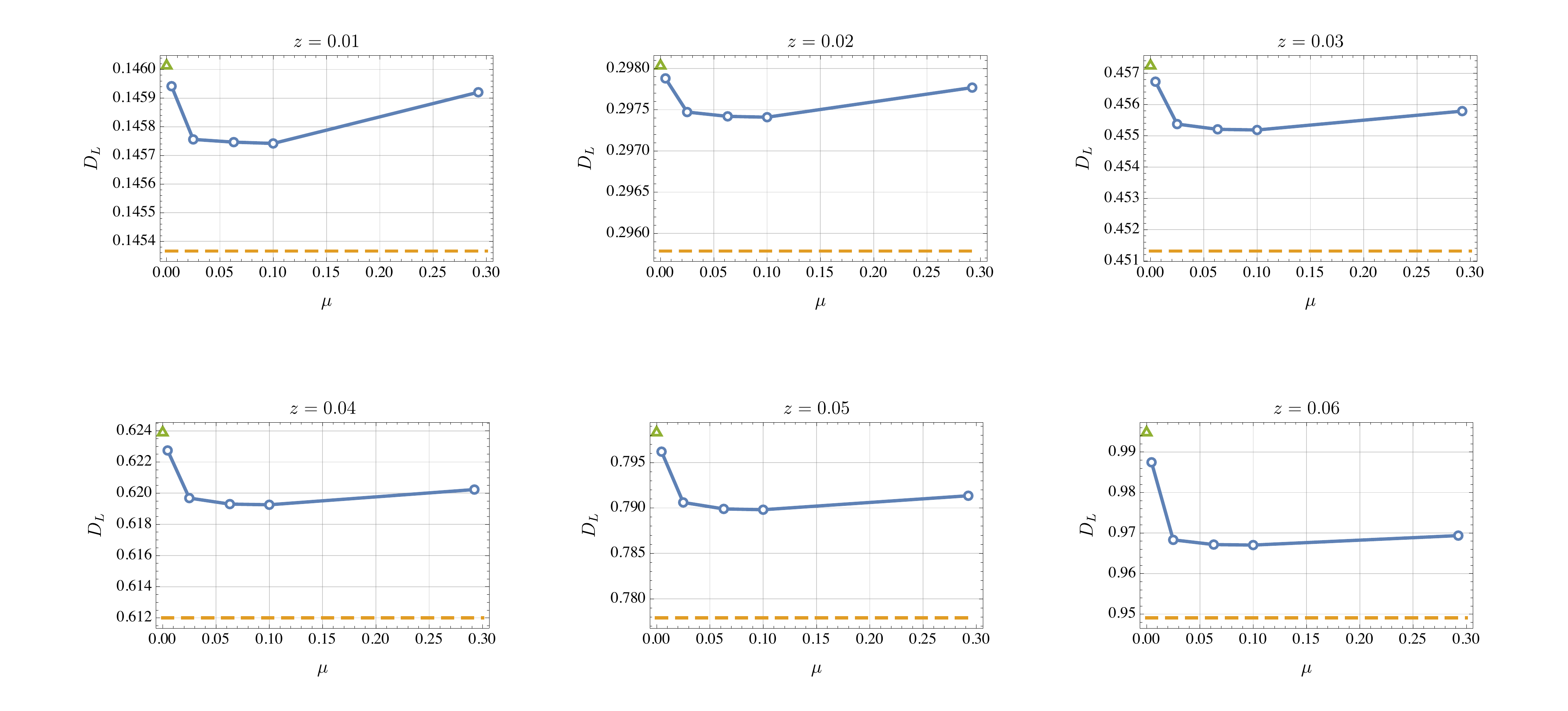}
\caption{
Behaviour of the luminosity distance
at fixed redshift, for various values of $\mu$. The green triangle
represents the polynomial extrapolation of the data series for $\mu \to 0$,
while the yellow
dashed curve represents the expected luminosity distance in EdS for each
specific value of $z$. 
\label{fig:murow}}
\efi

An important remark is that we observe that the tensor modes discussed in 
Section~\ref{sec:results} intensify as $\mu \to 0$, affecting the smaller-$\mu$
BHLs to the point that it becomes impossible to identify a monotonic trend
in the luminosity distance for large $z$.
For this reason, we are forced to limit our study to very small $z$.

\section{Discussion and conclusions}
\label{sec:dc}
We have investigated the propagation of light along two special curves in 
the spacetime of a BHL, constructed by numerically integrating Einstein's
equation in 3+1 dimensions. 
In particular, we have measured the redshift and luminosity distance along 
these curves, and compared them to the estimates of these observables
obtained in suitably fitted homogeneous models and in the EBA.
The comparison shows that the latter approximation is the one most capable
of reproducing the exact behaviour; we have built a heuristic argument to
explain this finding, based on the analysis of the different curvature
terms in the GDE.
Our finding is congruous with the conclusions of similar studies in other
inhomogeneous spacetimes~\cite{Fleury:2014gha}; in our case, however, the
models are not backreaction-free by construction, so that we can measure 
all the relevant contributions to the GDE. 

We have also fitted the $D_{\rm L}(z)$ relationship from the FLRW models with both a 
constant and a $z$-dependent $\Lambda$ to the data, finding that a value 
of $\Omega_\Lambda$ approximately equal to $\Omega_M$ reproduces the optical
properties of the BHL better than the corresponding models with $\Omega_\Lambda=0$.
In other words, in the BHL spacetime the luminosity distance for a redshift $z$
is larger than in the corresponding EdS model (the correspondence being based
on the same initial proper size and expansion rate). This is also in line
with the conclusions of previous studies~\cite{Clifton:2009bp}, and arguably
equivalent to the finding that fitting $\Omega_M$ alternatively leads to a smaller
value for this parameter~\cite{Fleury:2013uqa}.

Finally, we have examined a family of BHLs with varying BH masses and separations,
in order to estimate how our result changes as $\mu=M/L \to 0$. In this limit,
it was proven in~\cite{Korzynski:2013tea} that the expansion history of
a BHL tends to that of a flat FLRW model with the same average density.
Here, however, we find that the optical properties of a BHL exhibit a finite
deviation from the corresponding FLRW model, which reaches $5\%$ by $z=0.06$. 
Given a considerable pollution by tensor modes, which we conjecture originate
in our initial-data construction, the luminosity distance is oscillatory, and 
we are unable to evaluate the continuum limit for larger $z$.

Building a picture of the mechanisms involved in these results, as well as 
generalizing it to inhomogeneous spacetimes with different matter content
and density profiles, is a particularly intriguing but hard-to-approach task.
We can start to tackle it by comparing our results to a recent study~\cite{Giblin:2016mjp}, 
which also measured the effects of light propagation in an inhomogeneous 
model which, unlike the ones considered in this work, was filled with dust.
In that investigation, percent-level deviations were detected from the 
homogeneous Hubble law, which are about an order of magnitude smaller than 
the deviations reported here. From the arguments presented in this paper,
we infer that the discrepancy is largely due to the different representation
of the matter filling the two models. The quantitative formulation of 
this statement is a problem which we reserve for further study.

\section*{Acknowledgements}
MK and EB would like to thank the Max Planck Institute for Gravitational 
Physics (Albert Einstein Institute) in Potsdam for hospitality. The work 
was supported by the project \emph{``The role of small-scale inhomogeneities 
in general relativity and cosmology''} (HOMING PLUS/2012-5/4), realized 
within the Homing Plus programme of Foundation for Polish Science, 
co-financed by the European Union from the Regional Development Fund, and by 
the project ``\emph{Digitizing the universe: precision modelling for precision 
cosmology}'', funded by the Italian Ministry of Education, University and Research (MIUR).  
Some of the computations were performed on the Marconi cluster at CINECA.

\appendix
\section{Geodesic equation 3+1 decomposition}
\label{app:geo3+1}

The tangent $p^a$ to a geodesic satisfies

  \begin{equation}
    p^a \nabla_a p^b = 0 \\
  \end{equation}
  or, to simplify the following derivation,
  \begin{equation}
    p^a \nabla_a p_b = 0. \\
  \end{equation}
  The covariant derivative is expanded in terms of the partial
  derivative and the Christoffel symbol of the spacetime metric,
  \begin{equation}
    p^a \partial_a p_b = p^a p^c \Gamma_{cab} \\
  \end{equation}
  and the LHS is recognised as the derivative along the curve of the component $p_b$
  with respect to the curve parameter,
  \begin{equation}
    \frac{dp_b}{d\lambda} = p^a \partial_a p_b
  \end{equation}
  The Christoffel symbol is expressed in terms of derivatives of the metric,
  \begin{equation}
    p^a p^c \Gamma_{cab} = \frac{1}{2} p^a p^c(g_{cb,a} + g_{ac,b} - g_{ab,c})
  \end{equation}
  and we note that the first and third terms in parentheses are
  antisymmetric in $a$ and $c$, whereas $p^a p^c$ is symmetric, so
  these terms sum to zero, giving
  \begin{equation}
    \frac{dp_b}{d\lambda} = \frac{1}{2} p^a p^c g_{ac,b}, \label{eq:geoode4}
  \end{equation}
  as the geodesic equation for the covariant component of the tangent
  vector.

  We now wish to express the RHS in terms of the 3+1 quantities
  available in a numerical relativity simulation.  We summarise
  briefly the standard 3+1 decomposition of a spacetime (see, e.g.~\cite{alcubierre2008introduction}).
  A foliation of constant-time hypersurfaces is represented by a
  one-form $\Omega_a$, which locally can be written as the
  differential of the coordinate time, $\Omega_a = \nabla_a t$.  The
  lapse function is defined as $\alpha \equiv (-\Omega_a
  \Omega^a)^{-1/2}$, and the timelike unit hypersurface normal as $n_a \equiv -
  \alpha \Omega_a$, so that $n_a n^a = -1$.  The spatial metric on the
  hypersurface is
  \begin{equation}
  \gamma_{ab} = g_{ab} + n_a n_b \label{eq:g3+1}
  \end{equation}
  with $\gamma_{ab} n^a = 0$.  A vector $S^a$ is described as spatial
  if $S^a n_a = 0$.  Since the direction of time evolution, $t^a =
  (\partial/\partial t)^a$, is not necessarily aligned with the normal
  to the hypersurface, we express it in terms of a normal component,
  and the spatial {\em shift vector} $\beta^a$,
  \begin{equation}
    t^a = \alpha n^a + \beta^a. \label{eqn:tinnbeta}
  \end{equation}
  To simplify the expressions, we will work in the standard coordinate
  basis in which the timelike basis vector is $t^a \equiv
  (\partial/\partial t)^a$, and hence has components $[1,0,0,0]$.  In
  such a basis, we have $n_0 = -\alpha$, $n_i = 0$, $n^0 =
  \alpha^{-1}$, $S^0 = 0$, where $S^a$ is any spatial vector, and lower case Latin indices from the middle of the alphabet ($i,j,\ldots$) indicate
  spatial components (i.e.~$i = 1, 2, 3$).  These
  relations will simplify the derivation.

  We now need to express equation (\ref{eq:geoode4}) in terms of partial
  derivatives of the spatial quantities available in an NR simulation.
  We first decompose $p_a$ into a timelike and spatial part,
  \begin{equation}
    p_a = \sigma n_a + q_a \label{eq:p3+1}
  \end{equation}
  where $q_a n^a = 0$, and aim to find an equation for the evolution
  of $q_i$.  Note that $p_i = q_i$ since $n_i$ = 0.  Substituting
  equations (\ref{eq:g3+1}) and (\ref{eq:p3+1}) into equation (\ref{eq:geoode4}), we
  eventually obtain
  \begin{equation}
    \frac{dq_i}{d\lambda} = \sigma^2 n^a n_{a,i} - \sigma n^a_{,i} q^c
    \gamma_{ac} + \frac{1}{2} q^a q^c \gamma_{ac,i} .
  \end{equation}
  In deriving this, we have made use of the fact that all contractions
  of $(n_a n_c),_i$ with $q^a$ vanish, $n_i = 0$, $n^a n^c
  \gamma_{ac,i} = -n^a_{,i} n^c \gamma_{ac} = 0$, since $n^c
  \gamma_{ac} = 0$, and that $n^a \gamma_{ac,i} = -n^a_{,i}
  \gamma_{ac}$.

  We now express $n$ in terms of $\alpha$ and $\beta$ using
  equation (\ref{eqn:tinnbeta}) to obtain
  \begin{equation}
    \frac{dq_i}{d\lambda} = -\sigma^a \alpha^{-1} \alpha_{,i} + \sigma
    \alpha^{-1} \beta^a_{,i} q_a -\frac{1}{2} q_a q_c \gamma^{ac}_{,i}
  \end{equation}
  where we have used the fact that $t^\mu = \delta^\mu_0$, hence
  $t^\mu_{,\nu} = 0$.

  The tangent vector components are related to the
  coordinates of the curve via $dx^\mu/d\lambda = p^\mu$.  The time
  component gives $dt/d\lambda = p^0 = \sigma \alpha^{-1}$ and the
  spatial components give $dx^i/d\lambda = p^i$.  Using the chain rule,
  we obtain $d/dt = (p^0)^{-1} d/d\lambda$, and eliminating $\sigma$ in
  favour of $p^0$, we obtain finally
  equations (\ref{eqn:geo3+1x})--(\ref{eqn:geo3+1lambda}):
  \begin{eqnarray}
    \frac{d x^i}{dt} &=& -\beta^i+(p^0)^{-1} \gamma^{ik} q_k \\
    \frac{dq_i}{dt} &=& -p^0 \alpha \alpha_{,i} + q_j \beta^k_{,i} \gamma_{kj} - \frac{1}{2} (p^0)^{-1} q_l q_m \gamma^{lm}_{,i} \\
    \frac{d \lambda}{dt} &=& (p^0)^{-1}
  \end{eqnarray}
  in agreement with \cite{Hughes:1994ea}.
  
\section{Geodesic integrator tests}
\label{app:geo}
We now present three different tests of the 3+1 \code{Geodesic} code against existing known solutions: 
(i) the redshift-luminosity relationship in an EdS universe,
(ii) the geodesics in the Schwarzschild spacetime, and (iii)
the geodesics in a model from the Szekeres class.

\subsection{Redshift and luminosity distance in the EdS spacetime}

In the first test, we check both the redshift and the luminosity
distance by letting geodesics propagate in an EdS universe.

Using our infrastructure, we can propagate null rays on an EdS
universe and compare the answer to the analytical solution
(\ref{eq:ldeds}). Specifically, we use the code presented in~\cite{Bentivegna:2016stg}
to evolve a periodic cubic domain of this spacetime, with 
boundaries $-L/2 \leq x,y,z \leq L/2$, with $L=20$, starting with the same initial 
scale factor $a_{\cal S}$ and expansion rate $H_{\cal S}$ as the 
$M=1$ BHL discussed in section~\ref{sec:results}. We perform three 
separate runs with $5^3$, $10^3$ and $20^3$ points, tracking a
geodesic that moves from the origin along the $x$ axis (notice
that as this type of space is completely homogeneous and isotropic, 
the curve does not depend on the location of the null rays, but 
merely on the value of the scale factor at its end points). 

Figure~\ref{fig:eds} illustrates our result: as expected, the 
numerical solution converges to the exact equation (\ref{eq:ldeds}) 
at fourth order. 

\bfi
\bce
\includegraphics[width=0.45\textwidth]{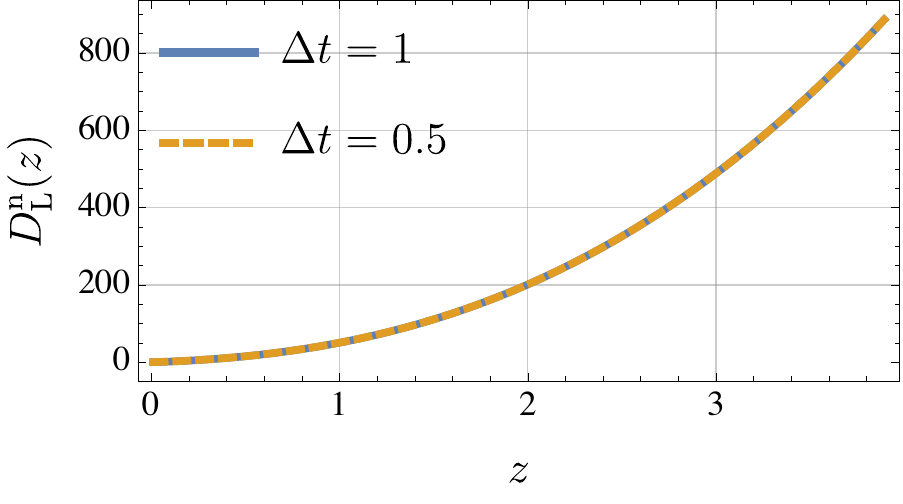}
\includegraphics[width=0.45\textwidth]{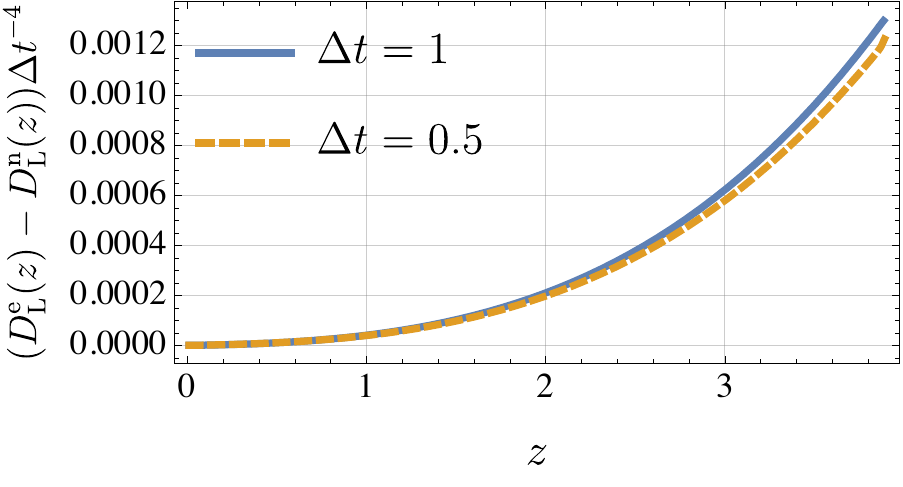}
\caption{Luminosity distance as a function of redshift for
the Einstein-de~Sitter model. Left: the numerical solution
$D^{\rm n}_L(z)$ for two different resolutions. Right:
the difference between the numerical and the exact solution
$D^{\rm e}_L(z)$ (given by (\ref{eq:ldeds})) multiplied
by $\Delta t^{-4}$ to show fourth-order convergence.
\label{fig:eds}}
\ece
\efi

\subsection{Geodesics in the Schwarzschild spacetime}

We then compare the computation of a geodesic in a numerical
Schwarzschild spacetime using the generic 3+1 \code{Geodesic} code
against a direct numerical integration of the well-known Schwarzschild
geodesic equations.

In Schwarzschild, the geodesic equation in the $\theta=\pi/2$ plane reduces to
\begin{eqnarray}
  \frac{dt}{d\lambda} &=& \frac{E}{1-2M/r} \label{geoscheqs1}\\
  \frac{d\phi}{d\lambda} &=& \frac{L}{r^2} \\
  \frac{d^2r}{d\lambda^2} &=& -\frac{M}{r^3}\left(r-2M\right)\left(\frac{E}{1-2M/r}\right)^2 + \frac{M}{r(r-2M)} \left(\frac{dr}{d\lambda}\right)^2 \nonumber \\
  && +(r-2M) \left ( \frac{L}{r^2} \right)^2 \label{geoscheqs3}
\end{eqnarray}
where $E$ and $L$ are the conserved energy and angular momentum (see,
for example, \cite{Carroll:2004st}).  Since there is no closed-form
solution, we integrate (\ref{geoscheqs1})--(\ref{geoscheqs3}) numerically
using Mathematica.

We then use the 3+1 \code{Geodesic} code to integrate the geodesic
using the same initial conditions on a uniform Cartesian grid of
spacing $\Delta x$ with timestep $\Delta t = 2 \Delta x$.  The metric
is expressed in isotropic coordinates, where the relation between
isotropic ($R$) and Schwarzschild ($r$) radial coordinates is
\begin{equation}
r = R (1 + M/(2 R))^2.
\end{equation}

The test null geodesic has initial conditions
\begin{eqnarray}
  [R,\theta,\phi] &=& [8M,0,0] \label{geoschic1}\\
  {[q_R,q_{\theta},q_{\phi}]} &=& [\sqrt{2},0,3\pi/4 ]\\
  \frac{d\lambda}{dt} &=& 1.  \label{geoschic2}
\end{eqnarray}

We compute the 3+1 solution using two
different grid spacings, $\Delta x = M/4$ and $M/8$, to assess
convergence of the solution.

\begin{figure}[h]
  \centering
  \begin{subfigure}[t]{0.9 \textwidth}
    \includegraphics[scale=0.7]{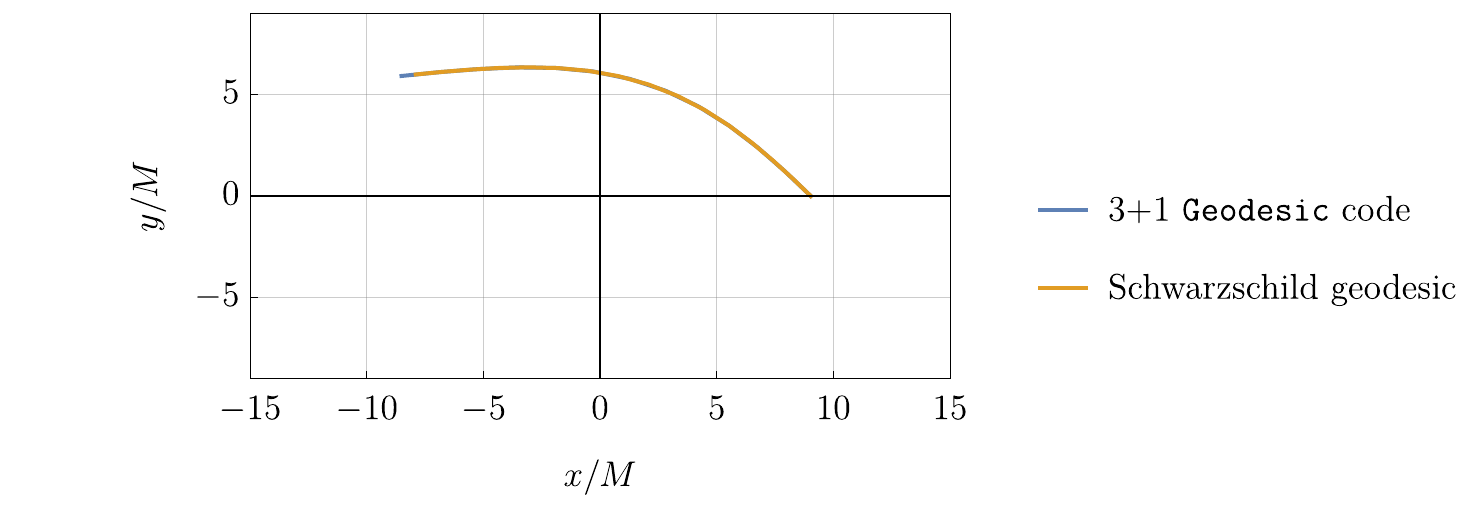}
  \end{subfigure}
  \\
  \begin{subfigure}[t]{0.9 \textwidth}
    \includegraphics[scale=0.7]{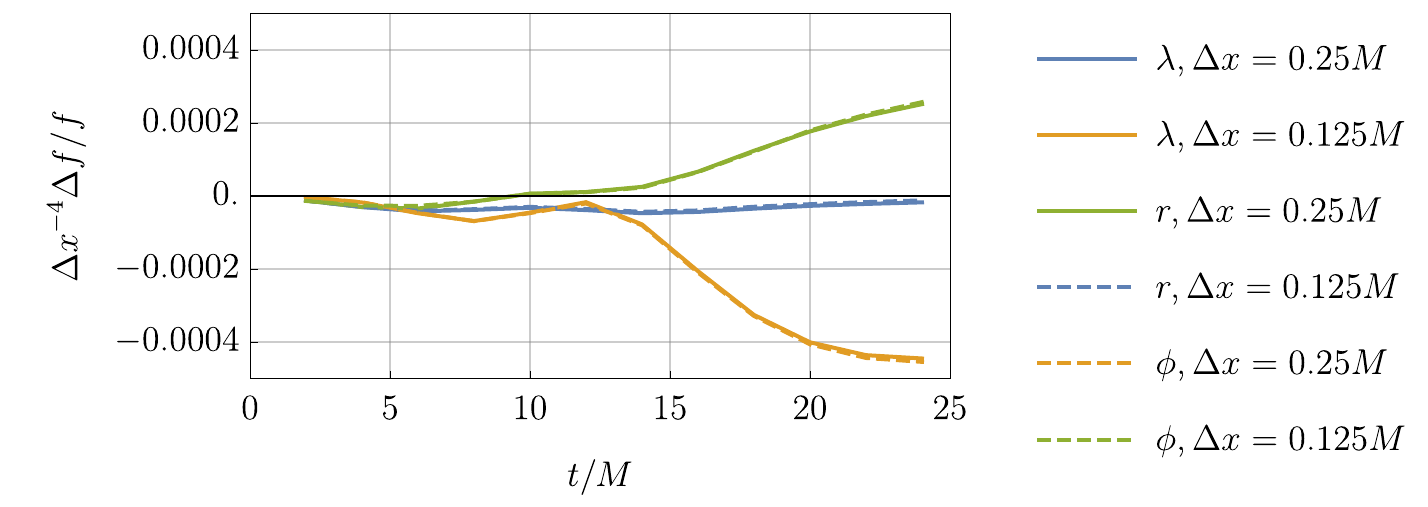}
  \end{subfigure}
  \caption{Test of the 3+1 \code{Geodesic} code against direct
    numerical integration of the Schwarzschild geodesic equations. Top: null geodesic trajectory in the $xy$ plane.
    Bottom: convergence of the affine parameter $\lambda$ and polar
      coordinates $r,\phi$ to the Schwarzschild geodesic solution.
      The errors have been rescaled by ${\Delta x}^{-4}$ and the
      agreement shows fourth order convergence.\label{fig:sch}}
\end{figure}

In Figure~\ref{fig:sch} we plot the trajectories of the Schwarzschild and 3+1 geodesics in
the $xy$ plane and see that they agree very well.  The numerical integration of the Schwarzschild
geodesic equations is much more accurate than the 3+1 solution, so we take the difference between the 3+1 and Schwarzschild solutions
to be the error in the 3+1 solution.
In Figure~\ref{fig:sch}, we also plot the error for
the three components of the solution, $\lambda(t)$, $r(t)$ and
$\phi(t)$ multiplied by ${\Delta x}^{-4}$ for the two different values
of $\Delta x$.  The curves agree well, indicating that the 3+1
\code{Geodesic} code produces a result which converges at fourth order to the
Schwarzschild geodesic solution,
\begin{equation}
  f_\mathrm{3+1} - f_\mathrm{Sch} = O({\Delta x}^4).
\end{equation}
The observed fourth order convergence is consistent with the expected
dominant error from the fourth-order Runge-Kutta time integrator.

\subsection{Geodesics in a spacetime from the Szekeres class}
A less trivial testbed is provided by the Szekeres class of inhomogeneous
cosmological models. We use, in particular, the axisymmetric solution
described by Meures and Bruni~\cite{Meures:2011gp,Meures:2011ke}, given
by the line element:
\beq
ds^2 = -dt^2 + S(t)^2 \left[ dx^2 + dy^2 + Z(t,z)^2 dz^2\right]
\eeq
with
\bea
&&S(t) = \left ( \frac{1-\Omega_\Lambda}{\Omega_\Lambda} \right)^{1/3} \sinh^{2/3}
\left [\frac{3}{2} H_0 \sqrt{\Omega_\Lambda} (t+t_\star) \right ] \\
&&Z(t,z) = 1 + (1-\sin kz )[f_+(t+t_\star)+B(x^2+y^2)] 
\eea
and $\Lambda$, $\Omega_\Lambda$, $k$ and $B$ are the cosmological constant and its associated density
parameter, an arbitrary wave number, and a constant given by:
\beq
B=\frac{3}{4} H_0^2 \left [ \Omega_\Lambda (1-\Omega_\Lambda)^2 \right ] ^{1/3},
\eeq
respectively.
Finally, $f_+(t)$ is a solution of:
\beq
f'' + \frac{4}{3} \coth \left (\sqrt{\frac{3 \Lambda}{4}} t \right ) f' - \frac{2}{3} \frac{1}{\sinh^2 \left (\sqrt{\frac{3 \Lambda}{4}} t \right )} f = 0.
\eeq
In~\cite{Meures:2011ke}, an ODE system for the geodesics propagating on 
this spacetime is provided. For geodesics 
propagating along the symmetry axis $x=y=0$, this system has the simplified 
form~\cite{Meures:2011ke}:
\bea
\label{eq:smbE}
&& -\frac{E'}{E}-\frac{S'}{S}-\frac{\dot F}{1+F} = 0 \\
\label{eq:smbz}
&& z'-\frac{2}{3} \frac{1}{H_0 \sqrt{\Omega_\Lambda} S Z} = 0
\eea
where $E$ is the photon energy along the geodesic and $z$ is its coordinate.
Primes indicate derivatives with respect to the rescaled time $\tau = \sqrt{3 \Lambda/4}t$. 
As in the Schwarzschild test, we solve this system with Mathematica.

As in~\cite{Bentivegna:2016stg}, we compute the metric of this spacetime 
on a cubic domain $[-L/2,L/2]^3$, with $L=2$, and at two different resolutions
$\Delta x={0.1,0.2}$.
Figure~\ref{fig:mb} shows the comparison between the numerical computations
and the solution of (\ref{eq:smbE})-(\ref{eq:smbz}). The agreement is
compatible with fourth-order convergence, as expected.

\bfi
\bce
\includegraphics[width=0.45\textwidth]{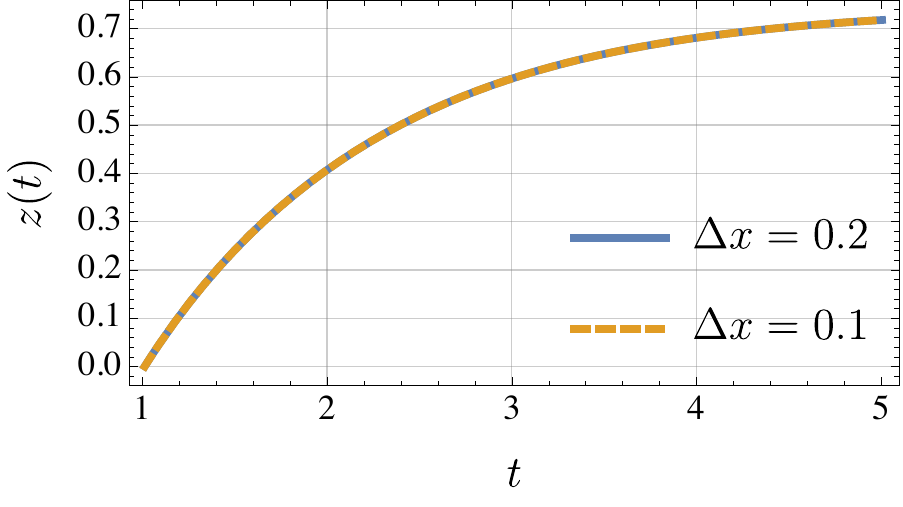}
\includegraphics[width=0.45\textwidth]{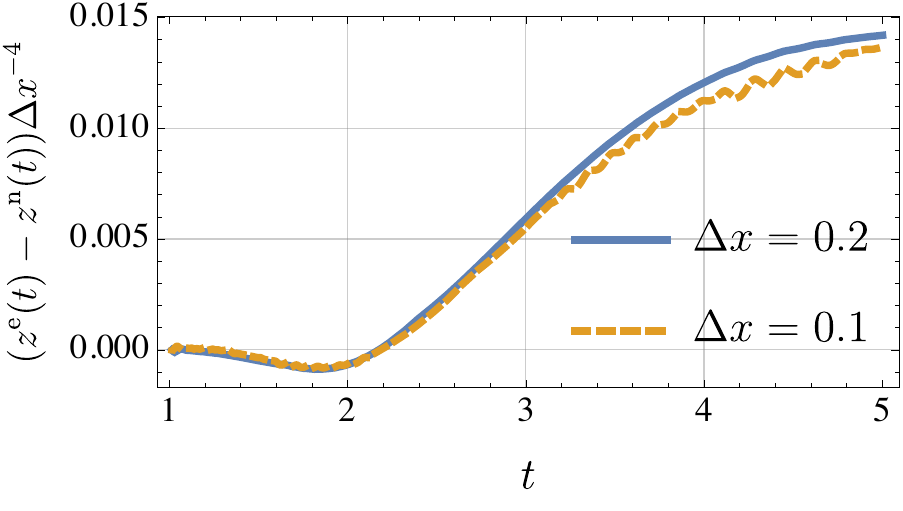}
\includegraphics[width=0.45\textwidth]{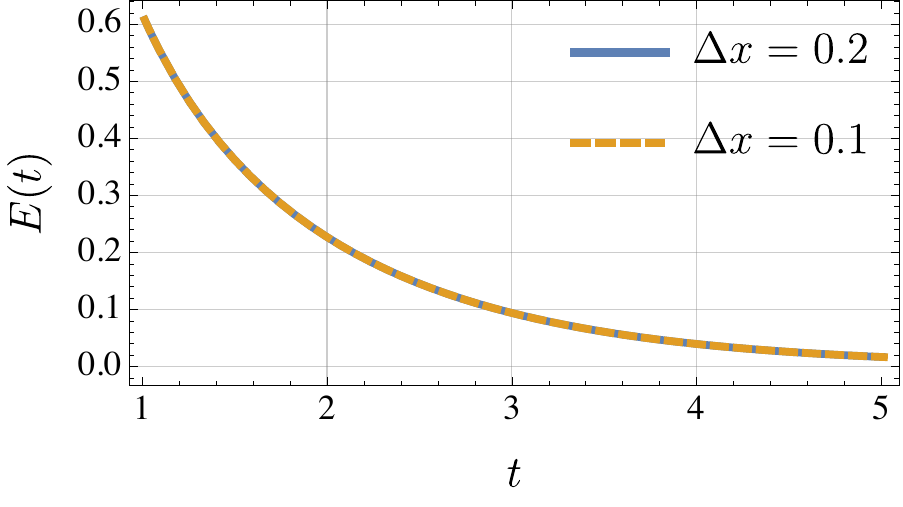}
\includegraphics[width=0.45\textwidth]{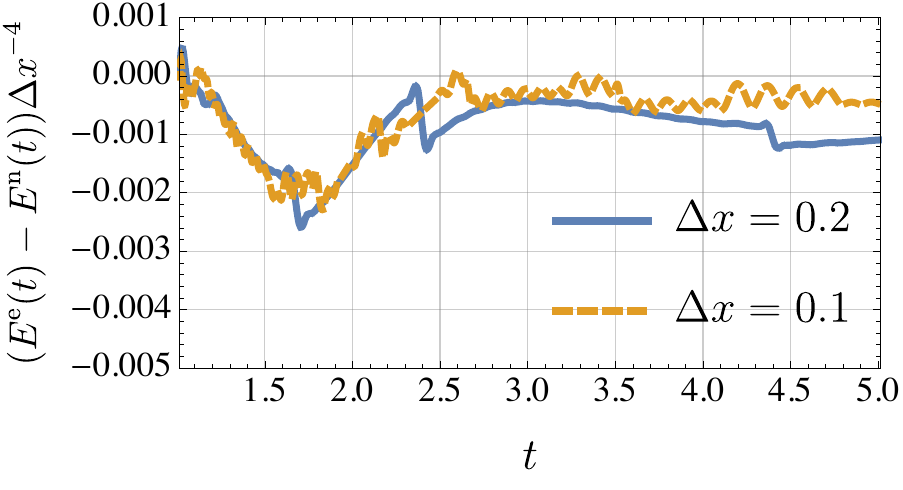}
\caption{Geodesic location and associated photon redshift
for a geodesic starting at the origin and parallel to the $z$ axis
in the Szekeres model.
\label{fig:mb}}
\ece
\efi

\bibliographystyle{jcap/JHEP}
\bibliography{references}

\end{document}